\documentclass[12pt,a4paper]{article}

\usepackage{amsmath,amsfonts,amssymb,amsthm,color}
\usepackage{mathtools}

\usepackage{latexsym}
\usepackage[pdftex]{graphicx}
\usepackage{epstopdf}

\usepackage{authblk}
\usepackage{dsfont}
\usepackage{amscd}

\usepackage{yfonts}

\usepackage{subfigure}

\usepackage{hyperref}

\usepackage[T1]{fontenc}

\def\nsection#1{\section{#1}\setcounter{equation}{0}}

\renewcommand{\theequation}{\thesection.\arabic{equation}}

\newcommand{\half}{\mbox{$\frac12$}}
\newcommand{\SL}{S^1}

\newcommand{\Diff}{\mathrm{Diff}}
\newcommand{\wDiff}{\widetilde{\mathrm{Diff}}}

\newcommand{\ee}{{\rm e}}
\newcommand{\ii}{{\rm i}}

\newcommand{\R}{\mathbb{R}}

\newcommand{\Z}{\mathbb{Z}}

\newcommand{\cD}{\mathcal{D}}
\newcommand{\cE}{\mathcal{E}}

\newcommand{\cG}{\mathcal{G}}
\newcommand{\cH}{\mathcal{H}}
 
\newcommand{\cJ}{\mathcal{J}} 

\newcommand{\cL}{\mathcal{L}}

\newcommand{\cO}{\mathcal{O}}
\newcommand{\cP}{\mathcal{P}}

\newcommand{\cR}{\mathcal{R}}

\newcommand{\cT}{\mathcal{T}}

\newcommand{\hc}{\mathrm{H.c.}}

\newcommand{\pdag}{^{\vphantom{\dagger}}}

\newcommand{\wick}[1]{\left. :\! \hspace{-0.5pt} #1 \hspace{-0.5pt} \!: \right.}

\def\sgn{\operatorname{sgn}}
\def\Tr{\operatorname{Tr}}

\newtheorem{theorem}{Theorem}[section]

\newtheorem{lemma}[theorem]{Lemma}

\theoremstyle{definition}

\theoremstyle{remark}

\title{%
Finite-time universality in nonequilibrium CFT
}%

\author[1]{%
Krzysztof Gaw\k{e}dzki%
\thanks{\,\texttt{kgawedzk@ens-lyon.fr}}%
}%

\author[2]{%
Edwin Langmann%
\thanks{\,\texttt{langmann@kth.se}}%
}%

\author[2]{%
Per Moosavi%
\thanks{\,\texttt{pmoosavi@kth.se}}%
}%

\affil[1]{%
Universit{\'e} de Lyon, ENS de Lyon,
Universit{\'e} Claude Bernard, CNRS,\break Laboratoire de
Physique, F-69342 Lyon, France \break
}%

\affil[2]{%
Department of Physics, KTH Royal Institute of Technology,\break
SE-10691 Stockholm, Sweden
}%

\date{%
\vspace{-0.8cm} \small July 18, 2018
}%

\begin{document}


\maketitle

\vskip -0.8cm

\hspace{8.5cm}{\it To Herbert, J\"urg, and Tom}
\vskip 0.9cm

\begin{abstract}
Recently, remarkably simple exact results were presented about the dynamics
of heat transport in the {\it local} Luttinger model for nonequilibrium
initial states defined by position-dependent temperature profiles. We present
mathematical details on how these results were obtained. We also give an
alternative derivation using only algebraic relations involving the
energy-momentum tensor which hold true in {\it any} unitary conformal field
theory (CFT). This establishes a simple universal correspondence between
initial temperature profiles and the resulting heat-wave propagation in CFT.
We extend these results to larger classes of nonequilibrium states. It is
proposed that such universal CFT relations provide benchmarks to identify
nonuniversal properties of nonequilibrium dynamics in other models. 
\end{abstract}


{%
\small
\noindent
{\bf Keywords:} Nonequilibrium dynamics -- Conformal field theory -- Heat and charge transport -- Luttinger model
}%


\nsection{Introduction}
\label{sec:Intro}
The study of heat, mass, charge, or spin transport in classical and quantum
one-dimensional systems has a long history, see, e.g., \cite{RLL,SpLe,ZNP,HoAr,Og1,AsPi,Gia,Zo,SPA1,SPA2}. Among problems that continue to make this an active field are questions concerning presence of diffusion, effects of integrability, interactions, or disorder, universality, and behaviors after quantum quenches, to mention a few. Studies of such questions were further spurred by experiments on ultracold atomic gases \cite{BDZ,PSSV} which recently triggered a rapid development of this field, see, e.g.,
\cite{CaCh,BCNF,CBT,BVKM2,DSpY,DSp,Spo,DoYo,IlDeN,CDDKY,Do}.
Let us specifically mention
the use of methods of conformal field theory (CFT) to gain better understanding of nonequilibrium steady states and transport in critical quantum $1d$ systems, see \cite{CaCa1,CaCa2,BeDo1,BeDo3} and references therein, \cite{HL}
for an operator-algebraic approach, and 
\cite{DSVC,BD1,DSC,BD2} that are particularly close to the context of
the present paper. 

In \cite{LLMM2} two of us (E.L. and P.M.) together with Joel L.
Lebowitz and Vieri Mastropietro studied in the Luttinger model the dynamics of heat transport starting from a particular class of nonequilibrium initial states. These states were given by position-dependent temperature profiles $1/\beta(x) > 0$, and the time evolution was determined by the standard
translation invariant Hamiltonian 
\begin{equation} 
\label{H} 
H = \int_{-L/2}^{L/2}\cE(x)dx,
\end{equation} 
where $\cE(x)$ is the {\it energy density} operator  
on a circle $\SL$ with circumference $L$ parameterized by the coordinate 
$x \in [-L/2, L/2]$. $\cE(x)$ together with the 
{\it heat current} operator $\cJ(x)$ satisfy the  continuity equation
\begin{equation} 
\partial_t\cE + \partial_x\cJ = 0
\end{equation}
with the usual Heisenberg time evolution $\cO(t) = \ee^{\ii Ht}\cO\ee^{-\ii Ht}$ for observables $\cO = \cE(x)$ and $\cJ(x)$.
(The units are such that $\hbar = k_B = e = 1$.)
We computed the evolution of the energy density and the heat current, $\langle\cE(x,t)\rangle_{\text{neq}}$ and $\langle\cJ(x,t)\rangle_{\text{neq}}$, using the following definition of nonequilibrium expectation values:\footnote{This definition differs from $\langle \cO \rangle$ in \cite{LLMM2} in that the thermodynamic limit is not taken in \eqref{cO1}.}
\begin{equation} 
\label{cO1} 
\big\langle\cO\big\rangle_{\text{neq}}
= \frac{\Tr(\ee^{-\cG}\cO)}{\Tr(\ee^{-\cG})}, 
\quad
\cG = \int_{-L/2}^{L/2}\beta(x)\cE(x) dx. 
\end{equation} 
Note that the special case of constant $\beta(x)=\beta_0$ corresponds to the standard Gibbs equilibrium expectations
\begin{equation} 
\label{Gibbs}
\big\langle\cO\big\rangle_{\beta_0}
=  \frac{\Tr(\ee^{-\beta_0H}\cO)}{\Tr(\ee^{-\beta_0H})} 
\end{equation}
at temperature $1/\beta_0$, and it is therefore natural to interpret $1/\beta(x)$ as a position-dependent temperature profile. We stick to this
interpretation throughout this paper, although a more common definition of
the local temperature would link it directly to
$\langle\cE(x,t)\rangle_{\text{neq}}$. Other
interpretations of $\beta(x)$ are possible, in particular if the states
in (\ref{cO1}) arise as equilibria for dynamics defined by inhomogeneous
Hamiltonians, as will be briefly discussed at the end of the paper.
The setup considered in \cite{LLMM2} resembled that of inhomogeneous 
quantum quenches \cite{SoCa} except that the evolution was studied after
quenches from {\it mixed} states of (\ref{cO1}) and it was analyzed directly
in real rather than imaginary time.

\begin{figure}[!htbp]
 
\centering

\subfigure[]{
  \includegraphics[scale=1.1, trim=0 0 0 0, clip=true]{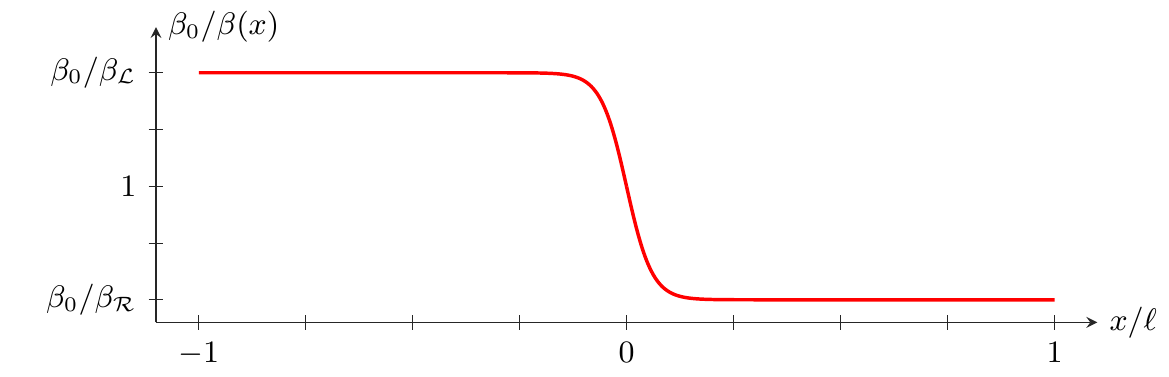}
\label{Fig:Luttinger_profile_subsystem}
}

\subfigure[]{
  \includegraphics[scale=1.1, trim=17 30 21 40, clip=true]{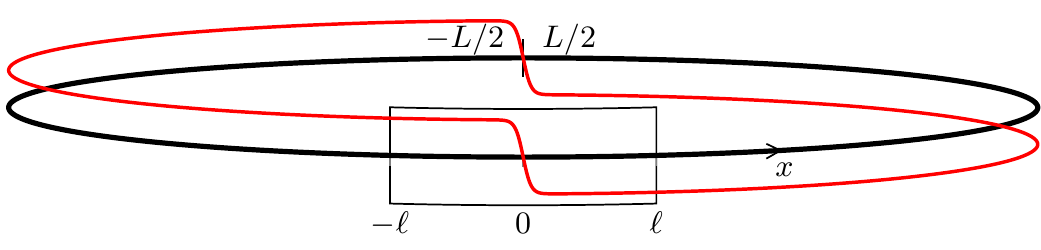}
\label{Fig:Luttinger_profile_PBC}
}

\caption{Temperature profile for \subref{Fig:Luttinger_profile_subsystem} the subsystem on a finite interval $[-\ell,\ell]$ with $L \gg \ell > 0$ and \subref{Fig:Luttinger_profile_PBC} the full system with periodic boundary conditions.
Note that, in addition to the kink at $x=0$, there is an opposite one at $x=\pm L/2$, which is necessary to have a smooth periodic function.
As explained in Sect.~\ref{sec:2}, the effect of this second kink is eliminated in the thermodynamic limit $L\to\infty$.}
\label{Fig:Luttinger_profile}
\end{figure}

\begin{figure}[!t]
\centering
\includegraphics[width=0.49\columnwidth, trim=10 0 3 0, clip=true]{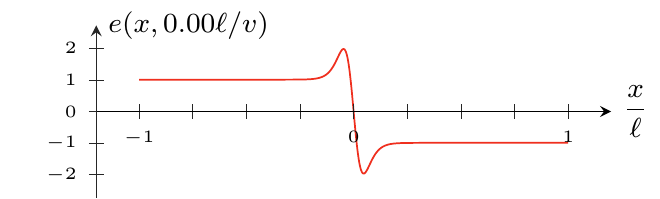}
\hfill
\includegraphics[width=0.49\columnwidth, trim=12 0 3 0, clip=true]{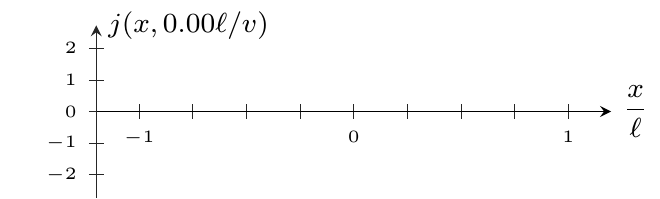}

\centering
\includegraphics[width=0.49\columnwidth, trim=12 0 3 0, clip=true]{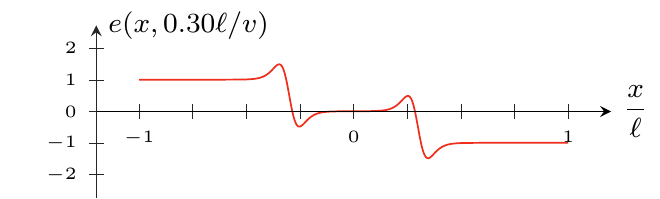}
\hfill
\includegraphics[width=0.49\columnwidth, trim=12 0 3 0, clip=true]{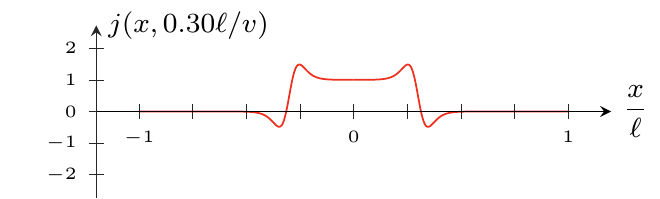}

\centering
\includegraphics[width=0.49\columnwidth, trim=12 0 3 0, clip=true]{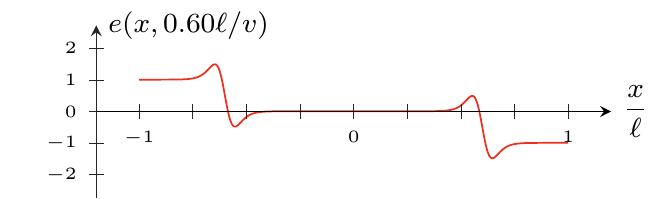}
\hfill
\includegraphics[width=0.49\columnwidth, trim=12 0 3 0, clip=true]{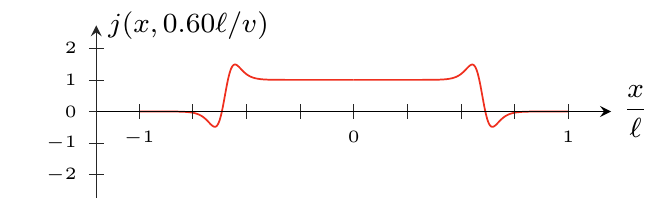}

\centering
\includegraphics[width=0.49\columnwidth, trim=12 0 3 0, clip=true]{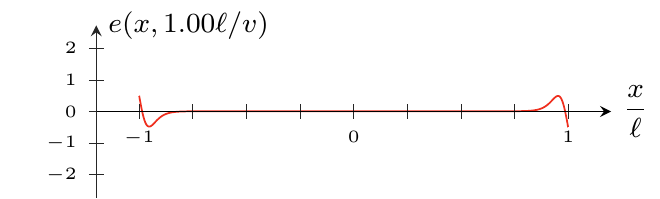}
\hfill
\includegraphics[width=0.49\columnwidth, trim=12 0 3 0, clip=true]{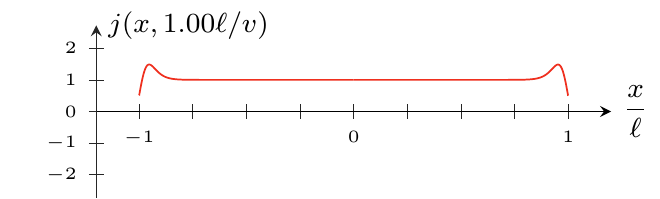}

\caption{Plots of the energy density $e(x,t) = v[\langle\cE(x,t) \rangle^\infty_{\text{neq}} -\langle\cE(x,t)\rangle^\infty_{\bar\beta}]/J$ and the heat current
$j(x,t) = \langle\cJ(x,t)\rangle^\infty_{\text{neq}}/J$ for the Luttinger model
in a finite interval $[-\ell,\ell]$ around $x = 0$ rescaled by
$J = \lim_{t \to \infty}\langle\cJ(x,t)\rangle^\infty_{\text{neq}} = ({\pi}/{12})
\left( \beta_{\cL}^{-2} - \beta_{\cR}^{-2} \right)$ for different times in
the nonequilibrium state with the inverse-temperature profile given by
\eqref{betaeps} and \eqref{Winfty}. The parameters are $\beta_{\cL} = 19.9$,
$\beta_{\cR}=20.1$, $\delta/\ell = 0.06$, and $v/\ell=0.025$.}
\label{Fig:Evolution_Luttinger}
\vskip 0.2cm
\end{figure}

To study heat transport we were particularly interested in kink-like profiles $1/\beta(x)$ interpolating between temperatures $1/\beta_{\cL}$ to the far left and $1/\beta_{\cR}$ to the far right, see Fig.~\ref{Fig:Luttinger_profile}.
The {\it smooth temperature profile protocol} described above allows one to analytically compute the nontrivial behavior of the energy density and the heat current around the location of the kink at early times and their subsequent development into {\it heat waves} moving ballistically to the right and left.
This should be contrasted with the results of the CFT description of the dynamics in the {\it partitioning protocol} employed in similar previous studies, see \cite{BeDo3} and references therein.
In such a description, argued to be valid after a transient time, the ballistic heat waves are compressed to simple jumps (shocks) without internal structure moving away from the contact point. In the smooth initial states that we
consider, this happens only in the limit when $t$ and $x$ are sent to infinity
at the same rate, as such a limit wipes out the nontrivial internal structure
of the heat waves. The evolution of the energy density and the heat current
obtained in \cite{LLMM2} permits then to better understand the shortcomings
of the partitioning protocol.
It also sheds a new interesting light on transport in integrable systems and, in particular, on how its universal features \cite{ViRi} emerge at long times
for a large class of nonequilibrium initial states, see \cite{LLMM1} and also
Sect.~\ref{subsec:4.3} below for a related discussion of charge transport
in the Luttinger model. 
As a representative example, we plot $\langle \cE(x,t) \rangle_{\text{neq}}$ and $\langle\cJ(x,t)\rangle_{\text{neq}}$ in Fig.~\ref{Fig:Evolution_Luttinger} at four times for the Luttinger model with local interactions (defined in more detail below) starting from the kink-like temperature profile in Fig.~\ref{Fig:Luttinger_profile}.
Note a peak and a dip in the energy density at time $t=0$ in the region where the temperature changes and how this local shape evolves into heat waves.
This is accompanied by a universal heat current building up in the region between the two heat-wave fronts.
We note that, for local interactions, the wave fronts preserve their shapes in time.
For the Luttinger model with nonlocal interactions, there are additional dispersive effects, which, however,
eventually become unobservable in any finite region as the wave fronts
leave such regions in finite time, see Fig.~1 in \cite{LLMM2}.
In the remainder of this paper we restrict our discussion to the local case.

In the {\it local Luttinger model}, the energy density is given
formally by\footnote{One can make this mathematically precise by considering
the analogous expression for a nonlocal interaction
and then taking the local limit \cite{ML} in an appropriate 
bosonized Fock space, see, e.g., \cite{LaMo}.}
\begin{equation} 
\label{cE}
\hspace*{-0.1cm}\cE(x)
= \hspace{-0.05cm}\sum_{r=\pm} \frac{v_F}{2}
		\! \wick{ \left[ \psi^\dag_r(x) (-\ii r\partial_x) \psi\pdag_r(x) +
    \hc \right] } \!
+\lambda\hspace{-0.14cm}\sum_{r,r'=\pm}\hspace{-0.18cm}
		\wick{ \psi^\dag_r(x)\psi\pdag_r(x) } \!
		\! \wick{ \psi^\dag_{r'}(x)\psi\pdag_{r'}(x)} \!
\,-\,\cE_0
\end{equation}
with fermionic field operators $\psi_{\pm}(x)$ (with antiperiodic boundary conditions) 
obeying the usual canonical anticommutation relations $\{\psi_r(x),\psi^\dag_{r'}(y)\} = \delta_{r,r'} \delta(x-y)$, etc., $\wick{\cdots}$ denoting Wick (normal) ordering, the bare Fermi velocity $v_F > 0$, and the coupling strength $\lambda > -\pi v_F/2$. The (diverging) constant $\cE_0$ subtracts the ground-state energy density
up to the finite contribution $-\pi v/(6L^2)$
left for later convenience.
The heat current is given by $\cJ(x)=v^2\cP(x)$ with the momentum density operator
\begin{equation} 
\label{momentumop}
\cP(x)
= \sum_{r=\pm} \frac{1}{2}
	\! \wick{ \left[ \psi^\dag_r(x) (-\ii\partial_x) \psi_r(x) + \hc \right] } \!
\end{equation}

\vskip -0.5cm
\noindent and the plasmon velocity $v = v_F \sqrt{1+2\lambda/(\pi v_F)}$ \cite{LLMM1,Voit}.
We found in \cite{LLMM2} the following exact expectation values of the
energy density and the heat current in the thermodynamic limit marked by
the superscript ${{}^\infty}$ on the expectations:\footnote{The symbols
$\cE$, $F$, and $f$ here correspond to $\cH$, $G$, and $g$ in \cite{LLMM2}.}
\begin{equation}
\label{EP}
\big\langle \cE(x,t) \big\rangle^\infty_{\text{neq}}
= \frac{1}{2} \left[ F(x^-)+F(x^+) \right],
\quad
\big\langle \cJ(x,t) \big\rangle^\infty_{\text{neq}}
= \frac{v}{2} \left[ F(x^-)-F(x^+) \right],
\end{equation} 
where $x^\pm = x\pm vt$ are the light-cone coordinates and the function
\begin{equation}
\label{E0}
F(x)
= \frac{\pi}{6v}\frac1{\beta(x)^2} + \frac{v}{12\pi}
	\left(
		\frac{\beta''(x)}{\beta(x)}
		- \frac{1}{2} \left( \frac{\beta'(x)}{\beta(x)} \right)^2
	\right) 
\end{equation}
is determined by the temperature profile.
We also observed that $F(x)$ can be written in terms of the Schwarzian derivative
\begin{equation} 
\label{S} 
(Sf)(x) = \frac{f'''(x)}{f'(x)} - \frac{3}{2} \left( \frac{f''(x)}{f'(x)} \right)^2
\end{equation}
of the function 
\begin{equation} 
\label{f}
f(x) = \int_0^x \frac{\beta_0}{\beta(x')} \, dx'
\end{equation}
as
\begin{equation} 
\label{F1} 
F(x) = \frac{\pi}{6v \beta_0^2}f'(x)^2 - \frac{v}{12\pi}(Sf)(x)
\end{equation}
for some constant $\beta_0>0$. 
Noting that Schwarzian derivatives appear in CFT \cite{FMS} and that the local Luttinger model is a CFT, we also argued that it should be possible to obtain the result in \eqref{EP} and \eqref{E0} in a simpler way using conformal transformations, and that it may be possible to also obtain expectation values of other observables in that way
\cite{LLMM2}.
Our main result in this paper is to show that this is indeed the case. 
\vskip 0.1cm

The method used in \cite{LLMM2} was perturbative in a small parameter $\epsilon$ measuring the distance to equilibrium in the initial state (i.e., the case $\epsilon=0$ corresponds to the Gibbs state).
This method is general, but generically one can only obtain useful low-order results.
In the special case of {\it local interactions}, however, we were able to push the computations to all orders in $\epsilon$, and, summing the resulting infinite series, we obtained the results in \eqref{EP} and \eqref{E0}.
In this paper we give a simpler derivation of these results extending them to {\it all} unitary CFT models, including models at finite $L$, and to other observables.
\vskip 0.1cm

Our analysis is based on the Minkowskian version of CFT, and it
generalizes to other classes of nonequilibrium states.
One such generalization is \eqref{cO1} but where, in addition to
the temperature profile $1/\beta(x)$, we also allow for a
{\it ``velocity'' profile} $\nu(x)$ taking
\begin{equation} 
\label{cG2} 
\cG = \int_{-L/2}^{L/2} \beta(x) \left[ \cE(x) + \nu(x)\cP(x) \right] dx
\end{equation} 
for $|\nu(x)| < v$ (this condition ensures that $\cG$ is positive).
Note that for constant profiles this is a generalization \cite{Callen} of the Gibbs state in \eqref{Gibbs} where $\beta_0H$ is replaced by $\beta_0(H+\nu_0 P)$ with the momentum operator 
\begin{equation} 
P = \int_{-L/2}^{L/2} \cP(x) dx.
\end{equation} 
This may be also be viewed in CFT as a Lorentz boost of the ordinary Gibbs state \cite{BeDo3}.
As will be shown, the corresponding generalization of the result in \eqref{EP} is obtained by replacing $F(x^\mp)$ by $F_{\pm}(x^\mp)$ given by \eqref{E0} with the right-hand side multiplied by the central charge $c$ of CFT (which is equal to 1 for the local Luttinger model) and with $\beta(x)$ replaced by
\begin{equation} 
\label{F2} 
\beta_{\pm}(x) = \beta(x)[1 \pm \nu(x)/v].  
\end{equation} 
In particular, in the long-time limit
\begin{equation}
\label{15}
\lim_{t\to\infty}\langle \cE(x,t) \rangle^\infty_{\text{neq}}
= \frac{\pi c}{12v} \left( \beta_{+,\cL}^{-2}+\beta_{-,\cR}^{-2} \right),
\quad
\lim_{t\to\infty}\langle \cJ(x,t) \rangle^\infty_{\text{neq}}
= \frac{\pi c}{12} \left( \beta_{+,\cL}^{-2} - \beta_{-,\cR}^{-2} \right),
\end{equation}
where $\beta_{\pm,\cL}$ and $\beta_{\pm,\cR}$ are the asymptotic values of $\beta_{\pm}(x)$ to the left and to the right, respectively.\footnote{To avoid confusion, we stress that our subscripts $\pm$ refer
to right $(+)$ and left ($-$) movers.}

Finally, for CFTs with a double $\text{U(1)}$ current algebra (e.g., the Luttinger model itself), we may also handle chemical-potential profiles $\mu_{\pm}(x)$ in addition to temperature profiles $1/\beta_{\pm}(x)$, possibly different for right and left movers.
In this case, the functions $F_{\pm}(x)$ pick up an additional term proportional to $\mu_{\pm}(x)^2$.
This is the largest class of nonequilibrium states that we consider. They are defined as in 
\eqref{cO1} but with $\cG$ in \eqref{cG_most_general}, see Sect.~\ref{subsec:4.3} 
for details.

The above results show that in CFT there is a universal
relation between the initial temperature, velocity, and chemical-potential
profiles and the resulting heat- and density-wave propagation even
at finite times.\footnote{Here we use the term ``universal'' as referring
to the same form in different CFTs rather than to the independence of
the microscopic details of the models.}
We believe that this provides a useful benchmark for other models as follows.
Typically, finite-time results are model-dependent, and more universal
behavior is only obtained at long times \cite{BeDo3}.
As an example, we mention the nonlocal Luttinger model which exhibits finite-time dispersion effects depending on short-distance details of the interaction potential \cite{LLMM2}.
However, these effects have some qualitative features that are always present.
We postulate that {\it the CFT results have to be subtracted in order
to identify the effects that come from the microscopic details} in the
propagation of the heat or density waves emanating from the inhomogeneities
of the initial state. In addition, such a subtraction should allow
to identify the time scales when such model-dependent effects
are important and when not.
Moreover, it was argued in \cite{Spo} that integrable systems come in two kinds: those that are purely ballistic and those with a nonzero diffusive contribution. For heat transport, in particular, one way to view this is through the thermal conductivity in the frequency domain
\begin{equation}
\label{Drude_def}
{\rm Re}\,\kappa_{\text{th}}(\omega) = \pi D_{\text{th}} \delta(\omega) +
{\rm Re}\,\kappa^{\text{reg}}_{\text{th}}(\omega),
\end{equation}
where a nonzero thermal Drude weight $D_{\text{th}}$ indicates the presence of a ballistic contribution and a finite nonzero value of the regular part
${\rm Re}\,\kappa^{\text{reg}}_{\text{th}}(\omega)$ at $\omega=0$
signals the presence of a diffusive component in the heat 
transport \cite{SPA1}.
It is known that ``pure'' CFT captures the ballistic part \cite{BeDo3},
with ${\rm Re}\,\kappa^{\text{reg}}_{\text{th}}=0$. This also follows from our results.
The cause of dispersive and diffusive effects thus must come from short-distance details, randomness, or other relevant perturbations, see, e.g., \cite{BeDo4} for a recent discussion of that issue within CFT.

The plan of the rest of this paper is as follows.
In Sect.~\ref{sec:2}, we sketch the original derivation of the result in \eqref{EP} and \eqref{E0} and explain the physical significance of the limit $L\to\infty$ since this is also relevant for our more direct CFT argument.
We also present special integrals whose exact evaluation was the key to this result and which, as we believe, are interesting in their own right.
The reader may skip the second half of Sect. 2 without
loss of continuity. The CFT derivation is given in Sect.~\ref{sec:3}.
After collecting the results about Minkowski-space CFT that are needed in Sect.~\ref{subsec:3.1}, we show in Sect.~\ref{subsec:3.2} how
to use conformal transformations to straighten out position-dependent temperature profiles $1/\beta(x)$ on a periodic interval.
This allows one to exactly map the nonequilibrium expectations in CFT to the corresponding equilibrium ones.
We make this mapping explicit for products of the components of the energy-momentum tensor.
In Sect.~\ref{subsec:3.3}, we study the thermodynamic limit of the finite-volume relations which allows one to treat temperature profiles on the infinite line with different asymptotic values on the left and right sides. As a
byproduct, we calculate the thermal Drude weight.   
Section~\ref{sec:4} is devoted to various generalizations.
In Sect.~\ref{subsec:4.1}, we briefly discuss the correlators of primary fields. 
In Sect.~\ref{subsec:4.2}, we consider states with different temperature
profiles for the right and left movers.
They form a class of nonequilibrium states preserved by the Schr\"odinger-picture dynamics that lead to simple examples of generalized Gibbs states
at long times.
In Sect.~\ref{subsec:4.3}, we extend the analysis to CFTs with a $\text{U(1)}$ current algebra and states with temperature and chemical-potential profiles. For the Luttinger model we discuss how this implies universality of conductance for both the charge and axial currents, generalizing previous results in \cite{LLMM1}, see also \cite{MW}. Finally, in Sect.~\ref{sec:5},
we make contact with \cite{DSVC,BD1,DSC,BD2}, discussing the dynamics
preserving states in \eqref{cO1} and the related Euclidian CFT
description. We end with conclusions and directions for further developments
in Sect.~\ref{sec:6}.
The \nameref{Sec:Appendix} contains some mathematical details on the special integrals mentioned above.


\nsection{Perturbative derivation and remarkable integrals} 
\label{sec:2}

The perturbative computation method used in \cite{LLMM2} is based
on introducing an expansion parameter $\epsilon$ measuring the deviation
of the temperature profile from constant temperature $1/\bar\beta$ as follows:
\begin{equation} 
\label{betaeps} 
\beta(x) = \bar\beta \left[ 1+\epsilon W(x) \right]
\end{equation}
with $W(x)$ a function defined by this relation. 

In \cite{LLMM2} we were mainly interested in kink-like functions where $W(x)$ becomes (say) $1/2$ and $-1/2$ to the far left and right, respectively, so that
$\bar\beta=(\beta_\cL+\beta_\cR)/2$ and
$\epsilon=(\beta_\cL-\beta_\cR)/\bar\beta$ in terms of the asymptotic values. 
On the other hand, for technical reasons, we used a model on a circle with
circumference 
$L<\infty$, which at first sight seems incompatible: a smooth function $W(x)$ on the circle with a single kink is not possible, and there has to be at least one other opposite one.
As an example, consider the periodic function
\begin{equation}
\label{Wexample}
W(x)
= - \frac{1}{2} \tanh \left(
		\frac{L}{2\pi\delta} \sin \left( \frac{2\pi x}{L} \right)
	\right)
\quad
(\delta>0),
\end{equation}
which is kink-like in the vicinity of $x=0$ as desired but has an opposite kink in the vicinity of $x=\pm L/2$ and leads (for negative $\epsilon$) to the
temperature profile of Fig.~\ref{Fig:Luttinger_profile}.
The effect of the additional step around $\pm L/2$ can be eliminated by computing results $\langle\cO(t)\rangle_{\text{neq}}$ for $L < \infty$ at finite times $t$, and then taking the limit $L\to\infty$ \cite{LLMM2} in which \eqref{Wexample} turns into
\begin{equation}
\label{Winfty}
W(x)= -\frac{1}{2} \tanh \left(\frac{x}{\delta}\right).
\end{equation}
In this way technical problems with $L=\infty$ on the level of quantum field theory are avoided and, at the same time, boundary conditions have no influence on the final results.
The physical interpretation is that the kink at $x = \pm L/2$ is ``behind the moon" and does not affect the physics in any finite region at times significantly smaller than $L/v$, which is a time scale that becomes infinite in the limit $L\to\infty$. 

In the rest of this section we sketch the perturbative derivation of the result in \eqref{EP} and \eqref{E0}, concentrating on remarkable integrals which were the key to this results.
The readers mainly interested in our CFT derivation may pass 
directly to Sect.~\ref{sec:3} without loss of continuity.  

With the inverse-temperature profile \eqref{betaeps}, one can use the Dyson
series to obtain an expansion
\begin{equation} 
\label{cOeps}
\langle\cO(x,t)\rangle_{\text{neq}}
= \langle \cO(x,t)\rangle_{\bar\beta}
	+ \epsilon \langle\cO(x)\rangle_1 + \epsilon^2 \langle\cO(x,t)\rangle_2 + \ldots 
\end{equation}
for any local observable $\cO(x,t)$, such as $\cE(x,t)$ and $\cJ(x,t)$.
The leading term in \eqref{cOeps} is the equilibrium expectation value (which is time independent, i.e., $\langle \cO(x,t)\rangle_{\bar\beta} = \langle \cO(x)\rangle_{\bar\beta}$), and the $n$-th order term is an $(n+1)$-point correlation function for $n=1,2,\ldots$ \cite{LLMM2}.
This method works, in principle, for {\it any} model but, in practice, it is difficult to go beyond leading order $n=1$.
For a quasi-free bosonic model, to which the Luttinger model reduces, one can use general mathematical results \cite{R,R1,GL} to replace the many-body computation by a much simpler one-particle one, and this makes it possible to extend the calculation to all orders \cite{LLMM2}. 

In particular, for the local Luttinger model, after taking the limit $L\to\infty$, this method gives \eqref{EP} with 
\begin{equation} 
\label{Fsum}
F(x) = \frac{\pi}{6v{\bar\beta}^{2}} + \sum_{n=1}^\infty \epsilon^n F_n(x),\quad 
\end{equation}
where
\begin{align} 
\label{Jn}
F_n(x)
& = \frac{v}{4\pi}\int_{\R^n}
		I_n(q_1,\ldots,q_n) \left( \prod_{j=1}^n\hat W(q_j)\ee^{\ii q_jx} \right)
		\frac{d^nq}{(2\pi)^n}, \\
\label{Jn'}
I_n(q_1,\ldots,q_n)
& = \frac{2}{v\bar\beta} \int_{\R}
		\sum_{\nu \in (2\pi/\bar\beta) \Z}
		\left( \prod_{j=0}^{n} \frac{v(p+Q_j)}{\ii\nu-v(p+Q_j)} \right) dp, \\
\label{Jn''}
\hat W(q)
& = \int_{\R} W(x) \ee^{-\ii qx} \, dx,
		\quad
		Q_j = \sum_{k=j+1}^n q_k.
\end{align}
It is interesting to note that the limit $L\to\infty$ is not only useful to eliminate the effect of boundary conditions (as explained) but also computationally: this limit turns Riemann sums into integrals and eliminates zero-mode contributions which would be more difficult to handle \cite{LLMM2}. 

The integrals in \eqref{Jn'} are certainly nontrivial, but we found that they all can be computed exactly, giving the result in Lemma~\ref{lemma:In}, and this was the key that led to \eqref{E0}. 

\begin{lemma}
\label{lemma:In} 
For all $n=1,2,\ldots$, 
\begin{multline}
\label{Jn4}
I_n(q_1,\ldots,q_n) \\
= (-1)^{n} \Biggl\{ \frac{n+1}{6} \left( \frac{2\pi}{v\bar\beta} \right)^2
	+ \frac{2}{(n+1)(n+2)} \sum_{j=0}^{n} Q_j^2
	- \frac{4}{n(n+1)(n+2)} \sum_{0 \leq j< k \leq n} Q_j Q_k \Biggr\}
\end{multline}
with $Q_j$ defined in \eqref{Jn''}.
\end{lemma}

\noindent A proof can be found in the \nameref{Sec:Appendix}.

It is remarkable that the result are even second order polynomials in the variables $q_j$.
It is clear from \eqref{Jn} that the constant term leads to contributions to $F_n(x)$ which are proportional to $W(x)^n$, whereas the terms with $q_j^2$ and $q_jq_k$, $j\neq k$, lead to $W(x)^{n-2}W''(x)$ and $W(x)^{n-2}W'(x)^2$, respectively.
Thus, the special form of the integrals in Lemma~\ref{lemma:In} implies that we get at most terms involving second derivatives of $W(x)$.
The explicit expression of these integrals allows one to compute $F_n(x)$ exactly, and the result is simple enough to analytically sum the series in \eqref{Fsum}, which gives the result in \eqref{E0} \cite{LLMM2}. 

We describe this computation above since it allows one to interpret the argument in the next section as a partial proof of Lemma~\ref{lemma:In}.
Such a proof is only partial since all terms with $q_j^2$ are identified with (say) $q_1^2$, and all terms with $q_jq_k$ for $j\neq k$ are identified with $q_1q_2$.
This identification is also useful in order to explicitly derive \eqref{E0} from Lemma~\ref{lemma:In}, see Eq.\ (A4) in \cite{LLMM2}, which is implied by \eqref{Jn4}.
Thus, the exact integrals in Lemma~\ref{lemma:In} contain more information than the result in \eqref{E0}.
Since nontrivial integrals that can be computed exactly are rare and often not only have one but several applications in physics, we prove \eqref{Jn4} in this paper.
Moreover, since our derivation of \eqref{EP} and \eqref{E0} in the next section works even for finite $L$, it suggests interesting Riemann sum generalizations of the exact integrals in Lemma~\ref{lemma:In}.
We believe it would be interesting to work them out, but this is left for a future study. 

We finally mention that our argument in the next section allows one to interpret the computation described above as a derivation of the conformal anomaly in CFT by a direct computation, see \eqref{key1}.


\nsection{CFT derivation}
\label{sec:3}


\subsection{CFT in Minkowski space} 
\label{subsec:3.1}

We consider the Minkowskian version of a unitary two-dimensional CFT where space is a circle $\SL$ parameterized by the periodic coordinate $x$ with the basic range $-L/2\leq x\leq L/2$ and where $t\in\R$ is time.
We keep the propagation speed $v$ in our equations to clearly indicate
how it effects the (otherwise) universal law relating the temperature
profiles to the heat-wave dynamics.

The basic objects of such a CFT are the periodic light-cone components $T_{\pm}(x^\mp)=T_{\pm}(x^\mp+L)$ of the energy-momentum tensor\footnote{In the 
more standard notation for the energy-momentum tensor components in
light-cone coordinates, $T_+=T_{--}$, $T_-=T_{++}$ and $T_{+-}=0=T_{-+}$.} where,
as before, $x^\pm = x \pm vt$.
They are distributions with values in the self-adjoint operators on the Hilbert space of states of the theory that satisfy the equal-time commutation relations
\begin{align}
[T_{\pm}(x),T_{\pm}(y)]
&= \mp 2\ii \delta'(x-y)T_{\pm}(y) \pm \ii\delta(x-y) T_{\pm}'(y)
\pm \frac{c}{24\pi}\ii \delta'''(x-y),
\label{Vir1}\\
[T_{\pm}(x),T_{\mp}(y)]
&= 0,\label{Vir2}
\end{align}
where $\delta(x)$ stands for the $L$-periodized $\delta$-function. 
The real number $c$ is the central charge of the theory.
In terms of the Fourier modes,
\begin{equation}
T_{\pm}(x)
= \frac{2\pi}{L^2}\sum_{n=-\infty}^{\infty} \ee^{\pm\frac{2\pi\ii nx}{L}}
	\left( L^{\pm}_n - \frac{c}{24} \delta_{n,0} \right),
\end{equation}
the commutation relations in \eqref{Vir1} and \eqref{Vir2} reduce to those of
the Virasoro algebra,
\begin{equation}
[L^\pm_n,L^\pm_m] = (n-m)L^\pm_{n+m}+\frac{c}{12}(n^3-n)\delta_{n+m,0},\qquad
[L^\pm_n,L^\mp_m] = 0.
\end{equation}
Technically, we assume that the Hilbert space of the theory is a (possibly infinite) direct sum of unitary highest-weight representations of two commuting copies of the Virasoro algebra.
The local Luttinger model is an example with $c=1$ of such a theory where
\begin{equation}
\label{LutTpm}
T_{\pm}(x) = \pi \! \wick{ \tilde{\rho}_{\pm}(x)^2 } \! - \frac{\pi}{12L^2},
\quad
\tilde{\rho}_{\pm}(x) = \rho_{\pm}(x) \cosh\varphi - \rho_{\mp}(x) \sinh\varphi  
\end{equation}
with the fermion densities $\rho_{\pm}(x) = \wick{ \psi^\dag_{\pm}(x) \psi\pdag_{\pm}(x) }$ and $\tanh 2\varphi = -\lambda/(\pi v_F+\lambda)$.
A related quantity describing the interactions is the Luttinger parameter\footnote{Eq.\,\eqref{LutTpm} holds also for the local Luttinger model with two coupling
constants $g_2$ and $g_4$ \cite{Voit} with $K=\ee^{2\varphi}$ dependent
on $g_2$ and $g_4$.}
$K = \ee^{2\varphi}$.
The effective densities $\tilde{\rho}_{\pm}$ act in a direct sum of bosonic Fock spaces that contains the interacting vacuum $|\Psi\rangle$ and the Wick ordering in \eqref{LutTpm} is with respect to $|\Psi\rangle$ \cite{ML,LaMo}.
In the following arguments, the explicit form of the operators $T_{\pm}(x)$ is not used. 

Let $\wDiff_+(\SL)$ denote the covering group of the group
$\Diff_+(\SL)$ of orientation-preserving diffeomorphisms of
the circle. The elements
of $\wDiff_+(\SL)$ are represented by smooth functions
$x \mapsto f(x)$ on $\mathbb R$ such that $f(x+L)=f(x)+L$ and
$f'(x)>0$, with functions $f(x)$ and $f(x)+nL$ corresponding to
the same diffeomorphism in $\Diff_+(\SL)$. The operator-valued distributions
$T_{\pm}$ generate two commuting projective unitary representations
$U_{\pm}$ of $\wDiff_+(\SL)$ on the Hilbert space of the theory such
that for infinitesimal diffeomorphisms $f(x)=x+\varepsilon\zeta(x)$
one has
\begin{equation} 
U_{\pm}(f) = I \mp\ii \varepsilon \int_{-L/2}^{L/2}\zeta(x)T_{\pm}(x) \, dx + o(\varepsilon)
\end{equation}
and under the adjoint action
\begin{equation} 
\label{key1} 
U_{\pm}(f) T_{\pm}(x) U_{\pm}(f)^{-1} = f'(x)^2T_{\pm}(f(x)) - \frac{c}{24\pi}(Sf)(x)
\end{equation}
with the Schwarzian derivative $(Sf)(x)$ given by \eqref{S}.
This was proven in \cite{GW2} for the unitary highest-weight representations of the Virasoro algebra and carries over to the present context. 
The adjoint action of $U_{\pm}(f)$ preserves $T_{\mp}$.

The energy-momentum tensor determines the Hamiltonian,
\begin{equation} 
\label{Hdef1}
H = v\int_{-L/2}^{L/2} \left[ T_{+}(x) + T_{-}(x) \right] dx, 
\end{equation}
and under the Heisenberg picture evolution, $T_{\pm}(x,t) = T_{\pm}(x^\mp)$ as claimed above. The energy and momentum density operators 
\begin{equation}
\label{cEcP_Tpm}
\cE(x,t) = v \left[ T_{+}(x^-) + T_{-}(x^+) \right],
\quad
\cP(x,t) = T_{+}(x^-) - T_{-}(x^+)
\end{equation}
satisfy the continuity equations
\begin{equation}
\partial_t\cE + v^2\partial_x \cP = 0,
\quad \partial_t\cP+\partial_x\cE = 0.
\end{equation}
As we shall see, the finite-volume nonequilibrium expectation values in \eqref{cO1} for $\cO = \cE(x)$ and $\cJ(x) = v^2\cP(x)$ are well defined for any such CFT provided the inverse-temperature profile is periodic.  


\subsection{Relating  nonequilibrium to equilibrium expectations} 
\label{subsec:3.2} 

From the above, it is clear that the calculation of the time evolution of the nonequilibrium expectation values of the energy density and current operators is equivalent to computing $\langle T_{\pm}(x^\mp)\rangle_{\text{neq}}$.

Let us denote $U(f) = U_{+}(f)U_{-}(f)$ for $f\in\wDiff_+(\SL)$. 
The key observation is that it is possible to find an $f$ such that
\begin{equation} 
\label{cond} 
U(f)\cG U(f)^{-1} = \beta_0 H + const 
\end{equation}
for some constant $\beta_0 > 0$.
In order to see this, take the function $f$ given by \eqref{f} with the constant $\beta_0$ determined by
\begin{equation}
\label{beta0}
\frac{1}{\beta_0}=\frac{1}{L}\int_{-L/2}^{L/2}\frac{1}{\beta(x')} \, dx'.
\end{equation}
The above choice of $\beta_0$ ensures that $f(x+L)=f(x)+L$ and thus that $f$ defines an element in $\wDiff_+(\SL)$.
Using this function $f$, it follows from \eqref{cO1}, \eqref{key1}, and \eqref{cEcP_Tpm} that
\begin{multline}
U(f) \cG U(f)^{-1}
= v \int_{-L/2}^{L/2}\beta(x)
		\left[ U_{+}(f)T_{+}(x)U_{+}(f)^{-1} + U_{-}(f)T_{-}(x)U_{-}(f)^{-1} \right] dx \cr
= v \int_{-L/2}^{L/2} \beta(x)f'(x)^2 \left[ T_{+}(f(x)) + T_{-}(f(x)) \right] dx
		- \frac{cv}{12\pi} \int_{-L/2}^{L/2} \beta(x) (Sf)(x) \, dx.
\end{multline}
Upon using the relation $f'(x) = \beta_0/\beta(x)$ for the derivative of $f$, this reduces to
\begin{align}
U(f)\cG U(f)^{-1}
& = v\beta_0 \int_{-L/2}^{L/2}f'(x) \left[ T_{+}(f(x)) + T_{-}(f(x)) \right] dx
  	- \frac{cv}{12\pi} \int_{-L/2}^{L/2} \beta(x)(Sf)(x) \, dx \nonumber \\
& = v\beta_0 \int_{-L/2}^{L/2} \left[ T_{+}(y) + T_{-}(y) \right] dy
		- \frac{cv}{12\pi} \int_{-L/2}^{L/2} \beta(x) (Sf)(x) \, dx,
\end{align}
where the last equality follows by the change of variables $y = f(x)$.
This establishes \eqref{cond}.
In short, the conjugation with $U(f)$ straightens out the temperature profile.

From the definition in \eqref{cO1} and the relation in \eqref{cond}, we infer that
\begin{equation}
\big\langle\cO\big\rangle_{\text{neq}} =
\frac{\Tr(\ee^{-U(f)\cG U(f)^{-1}}U(f)\cO U(f)^{-1})}{\Tr(\ee^{-U(f)\cG U(f)^{-1}})}=
\big\langle U(f)\cO U(f)^{-1}\big\rangle_{\beta_0}.
\label{transl}
\end{equation}  
This translates the nonequilibrium expectations to the equilibrium ones
defined by \eqref{Gibbs}. For $x^{-r}= x-rvt$ and
\begin{equation}
\label{Tpms}  
\cO = \prod_{j}T_{r_j}(x_j^{-r_j})
\end{equation}  
(with noncoincident points), using again \eqref{key1}, we obtain the relation
\begin{equation}
\label{neqexp}  
\Big\langle \prod_{j} T_{r_j}(x_j^{-r_j}) \Big\rangle_{\text{neq}}
= \Big\langle
\prod_{j} \left( f'(x_j^{-r_j})^2T_{r_j}(f(x_j^{-r_j}))
- \frac{c}{24\pi}(Sf)(x_j^{-r_j}) \right)
	\Big\rangle_{\beta_0}
\end{equation}
for $f$ given by \eqref{f} and \eqref{beta0}.
In particular,
\begin{equation}
\label{exp1pt}  
\big\langle T_{\pm}(x^\mp)\big\rangle_{\text{neq}}
= f'(x^\mp)^2\big\langle T_{\pm}(f(x^\mp))\big\rangle_{\beta_0}
	- \frac{c}{24\pi} (Sf)(x^\mp).
\end{equation}
The Gibbs state is translation invariant so that the expectations $\big\langle T_{\pm}(y)\big\rangle_{\beta_0}$ do not depend on $y$, but they depend,
in general, on $v\beta_0$ and $L$.
By scaling, however, $(v\beta_0)^2 \big\langle T_{\pm}(y) \big\rangle_{\beta_0}$
depends only on $v\beta_0/L$ but in a way dependent on the representation
content of the CFT. As we shall see, what is universal, depending only on
the central charge, is the $L\to\infty$ limit of $(v\beta_0)^2
\big\langle T_{\pm}(y)\big\rangle_{\beta_0}$.


\subsection{Thermodynamic limit}
\label{subsec:3.3}

Let us consider the limit $L\to\infty$ of the nonequilibrium expectations in \eqref{neqexp}. For a large class of kink-like $\beta(x)$ profiles (not necessarily symmetric) with an antikink around $\pm L/2$,
\begin{equation}
\beta_0^{-1} = \frac{1}{2} \left( \beta_{\cL}^{-1}+\beta_{\cR}^{-1} \right) + O(L^{-1}),
\end{equation}
where $\beta_{\cL}$ and $\beta_{\cR}$ are the asymptotic values of the plateau to the left and to the right of the kink.\footnote{Note that the constant $\beta_0$ defined by \eqref{beta0} differs from $\bar\beta$ in \eqref{betaeps} by an $O(\epsilon^2)$ term.}
Similarly, for fixed $x$, $\beta(x)$ will stabilize up to $O(L^{-1})$ terms with any trace of the antikink gradually wiped out, and so does the function
$f$ given by \eqref{f}.
The question about the large-$L$ limit of the nonequilibrium expectations in \eqref{neqexp} then boils down to the one for the equilibrium expectations
$\big\langle \prod_{j} T_{r_j}(x_j) \big\rangle_{\beta_0}$,
where, by rescaling, $\beta_0$ may be set to its asymptotic value and the insertion points are allowed to have $O(L^{-1})$ variations.

In CFT the control of the thermodynamic limit of the equilibrium expectations is an easy exercise.
The thermal expectations like those mentioned above may be viewed as the ones in the Euclidean theory on the torus
$\SL\times\SL$ where the circles have circumferences $L$ and $v\beta_0$, respectively.
In a modular invariant CFT \cite{FMS}, they also have a dual representation
\begin{equation}
\label{dual}
\Big\langle \prod_{j} T_{r_j}(x_j) \Big\rangle_{\beta_0}  
= \Big\langle \cT \prod_{j} \bigl(-T_{r_j}(\ii r_jx_j)\bigr) \Big\rangle_{L/v}
\end{equation}
as the equilibrium expectations with inverse temperature $L/v$ in the theory on the circle with circumference $v\beta_0$.
The components of the energy-momentum tensor with complex arguments on the right-hand side are defined by
\begin{equation}
T_{\pm}(x\pm\ii v\tau)=\ee^{\tau H}T_{\pm}(x)\ee^{-\tau H}
\end{equation}
and $\cT$ orders $x_j$ increasingly from the right to the left.
The identity in \eqref{dual} comes from swapping the two circles that play a symmetric role in the Euclidean version of the theory. 
It still holds for the Luttinger model if the finite-volume theory corresponds to antiperiodic boundary conditions for the fermionic fields $\psi_{\pm}(x)$, as
we assumed, even if the corresponding CFT does not have full modular
invariance.

When $L\to\infty$, the right-hand side of \eqref{dual} tends to the vacuum expectations providing the dual representation of the equilibrium expectations in the thermodynamic limit:
\begin{equation}
\label{dual0}  
\Big\langle \prod_{j} T_{r_j}(x_j) \Big\rangle^\infty_{\beta_0}=
\bigl\langle0\bigl| \cT \prod_{j} \bigl( -T_{r_j}(\ii r_jx_j) \bigr) \bigr|0\bigr\rangle.
\end{equation}
Besides, such vacuum expectations are universal because they receive contribution only from the tensor product of the two vacuum highest-weight representations of the Virasoro algebra.
They factorize according to
\begin{equation}
\label{factor}  
\bigl\langle0\bigl| \cT \prod_{j} \bigl( -T_{r_j}(\ii r_jx_j) \bigr) \bigr|0\bigr\rangle
= \bigl\langle0\bigl|
		\cT\hspace{-0.1cm}\prod_{j \, : \, r_j=+}\hspace{-0.1cm}\bigl( -T_{+}(\ii x_j) \bigr)
	\bigr|0\bigr\rangle
	\bigl\langle0\bigl|
	\cT\hspace{-0.1cm} \prod_{j \, : \, r_j=-} \hspace{-0.1cm}
        \bigl( -T_{-}(-\ii x_j) \bigr)
	\bigr|0\bigr\rangle.
\end{equation}
In the theory on the circle with circumference $v\beta_0$,
\begin{align}
\bigl\langle0\bigl| T_{\pm}(\pm\ii x_1) \bigr|0\bigr\rangle
& = - \frac{\pi c}{12(v\beta_0)^2},
		\label{1pt} \\
\bigl\langle0\bigl|\cT\, T_{\pm}(\pm\ii x_1)T_{\pm}(\pm\ii x_2) \bigr|0\bigr\rangle
& = \left( \frac{\pi c}{12(v\beta_0)^2} \right)^2
		+ \frac{\pi^2c}{8(v\beta_0)^4\sinh^4 \bigl(\frac{\pi}{v\beta_0}(x_1-x_2)\bigr)},
		\label{2pt} \\
\bigl\langle0\bigl| \cT\, T_{\pm}(\pm\ii x_1)T_{\mp}(\mp\ii x_2) \bigr|0\bigr\rangle
& = \left( \frac{\pi c}{12(v\beta_0)^2} \right)^2.\label{2pt-}
\end{align}

We infer that the identity in \eqref{neqexp} holds in the thermodynamic limit for infinite-volume profiles $\beta(x)>0$ with arbitrary asymptotic values $\beta_{\cL}$ and $\beta_{\cR}$ and the function $f$ defined by \eqref{f}.
By scaling, the right-hand side is then independent of the choice of $\beta_0$.
In particular, \eqref{exp1pt} together with \eqref{1pt} show that in the limit
$L\to\infty$,
\begin{equation}
\label{exp1ptinf}
\bigl\langle T_{\pm}(x^\mp)\bigr\rangle^\infty_{\text{neq}}
= f'(x^\mp)^2\frac{\pi c}{12(v\beta_0)^2} - \frac{c}{24\pi} (Sf)(x^\mp)
= \frac{\pi c}{12(v\beta(x^\mp))^2} - \frac{c}{24\pi} (Sf)(x^\mp).
\end{equation}
This is equivalent to \eqref{EP} with $F$ given by the right-hand side of \eqref{E0} multiplied by the central charge $c$. 

The above result allows one to easily extract the value of the thermal Drude weight $D_{\text{th}}$, see \eqref{Drude_def},
which may be obtained \cite{DSp,VKM} from
\begin{equation}
\label{Drude}
D_{\text{th}}
= -\beta_0^2\,\lim\limits_{\beta_{\cL,\cR}\to\beta_0}\frac{1}{\Delta\beta}
	\lim_{t\to\infty} \frac{1}{t}
	\int \big\langle\cJ(x,t) \big\rangle^\infty_{\text{neq}} \, dx,
\end{equation}
where the nonequilibrium expectation is calculated for the inverse-temperature profile $\beta(x)$ interpolating between the asymptotic values $\beta_{\cL}$ and $\beta_{\cR}$ with $\Delta\beta=\beta_\cL-\beta_\cR$.
The factor $-\beta_0^2$ is there to relate $D_{\text{th}}$ to the response to temperature rather than inverse-temperature change.
The space integral receives the contribution from the region of length $\approx2vt$ between the two ballistically separating heat waves where the heat current takes the long-time
value $({\pi c}/{12})(\beta_{\cL}^{-2}-\beta_{\cR}^{-2})$.
This yields
\begin{equation}
D_{\text{th}} = \frac{\pi v c}{3\beta_0}
\end{equation}
which is proportional to the temperature $1/\beta_0$. The
result may be also obtained using the partitioning protocol that leads
to the same steady state. It agrees with the calculation of the thermal
conductivity by the Green-Kubo formula,
\begin{align}
\kappa_{\text{th}}(\omega)
& = \beta_0\int_0^\infty\ee^{\ii\omega t}\,dt\int_0^{\beta_0} d\tau
\int\big\langle\cJ(x,t)\,\cJ(0,i\tau)\big\rangle^\infty_{\beta_0}\,dx \nonumber \\
& = \frac{\pi^2c}{8\beta_0^3}\int_0^\infty\ee^{\ii\omega t}\,dt\int_0^{\beta_0} d\tau
\int\sum\limits_{r=\pm}\sinh^{-4} \biggl (\frac{\pi(x-rvt+r\ii v\tau)}{v\beta_0}
\biggr) \,dx \nonumber \\
& = \frac{\pi vc}{4\beta_0}\int_0^\infty\ee^{\ii\omega t}\,dt\int\cosh^{-4}(y)\,dy
	= \frac{\pi vc}{3\beta_0}\Bigl(\pi\delta(\omega)+\ii\,PV\frac{1}{\omega}\Bigr),
\end{align}
where we used \eqref{2pt} and \eqref{2pt-} to express the infinite-volume
2-point correlation function of the heat current. Note that the regular part
of $\,{\rm Re}\,\kappa_{\text{th}}(\omega)$ vanishes, confirming the absence of
diffusive heat transport in nonequilibrium CFT. 


\nsection{Generalizations}
\label{sec:4} 


\subsection{Other correlators}
\label{subsec:4.1}

The finite-volume relation in \eqref{transl} between the nonequilibrium and equilibrium expectations may be rendered explicit also for observables
\begin{equation}
\cO = \prod_{j} \Phi_j(x^-_j,x^+_j),
\end{equation}  
where $\Phi_j(x^-_j,x^+_j)$ are primary fields whose transformation laws under the $\wDiff_+(\SL)$ symmetry take the form
\begin{equation}
U_{+}(f)U_{-}(f)\Phi_j(x^-,x^+)U_{-}(f)^{-1}U_{+}(f)^{-1} \\
= f'(x^-)^{\Delta^+_{\Phi_j}} f'(x^+)^{\Delta^-_{\Phi_j}}
	\Phi_j \left( f(x^-),f(x^+) \right)
\end{equation}
with the conformal weights $\Delta^\pm_{\Phi_j}\geq0$.
It then follows from \eqref{transl} that
\begin{equation}
\label{primneq}  
\Big\langle \prod_{j} \Phi_j(x^-_j,x^+_j) \Big\rangle_{\text{neq}}
= \Big\langle \prod_{j} \left( f'(x_j^-)^{\Delta^+_{\Phi_j}}f'(x_j^+)^{\Delta^-_{\Phi_j}}
	\Phi_j \left( f(x^-_j),f(x^+_j) \right) \right) \Big\rangle_{\beta_0}
\end{equation}
for $f$ given by \eqref{f} and $\beta_0$ by \eqref{beta0}.
Note that these are simpler relations than for the energy-momentum tensor components $T_{\pm}$ since those fail to be primary fields with conformal weights $(2,0)$ and $(0,2)$ due to the Schwarzian derivative term in \eqref{key1} reflecting the conformal anomaly.
In a similar way as for observables built from the operators $T_{\pm}$, the relations \eqref{primneq} hold also in the thermodynamic limit which may be controlled like before.
Also, as before, $\beta_0$ may be taken arbitrary in the infinite volume.  

For example, the Luttinger model has a conserved $\text{U(1)}$ current with the light-cone components $J_{\pm}(x^\mp) = \sqrt{K}\,\tilde{\rho}_{\pm}(x^\mp)$, see \eqref{LutTpm}, with conformal weights $(1,0)$ and $(0,1)$, respectively, and (renormalized) fermionic fields $\psi_{\pm}(x^-,x^+)$ with conformal weights
\begin{equation}
\left( \Delta^{+}_{\psi_{+}},\Delta^{-}_{\psi_{+}} \right)
= \left(\frac{(K+1)^2}{8K},\frac{(K-1)^2}{8K}\right),
\quad
\left( \Delta^{+}_{\psi_{-}},\Delta^{-}_{\psi_{-}} \right)
= \left(\frac{(K-1)^2}{8K},\frac{(K+1)^2}{8K}\right)
\end{equation}
accompanied by their Hermitian conjugates $\psi^\dagger_{\pm}(x^-,x^+)$ with the same conformal weights.\footnote{The fermionic fields are represented
as vertex operators related to the bosonic fields $\tilde\rho_\pm$.
Such operators require Wick ordering that provides their multiplicative
renormalization, see, e.g., \cite{LaMo}.}
Their infinite-volume equilibrium 2-point correlation functions have the form
\begin{equation}
\bigl\langle J_{\pm}(x^\mp)J_{\pm}(y^\mp) \bigr\rangle^\infty_{\beta_0}
= -\frac{K}{4 (v\beta_0)^2
  	\sinh^2\big(\frac{\pi}{v\beta_0}(x^\mp-y^\mp)\big)}
\end{equation}
and
\begin{multline}
\hspace*{1cm}\bigl\langle \psi_{\pm}(x^-,x^+)\psi^\dagger_{\pm}(y^-,y^+) 
\big\rangle^\infty_{\beta_0} \\
= \frac{
		\ee^{\pi \ii \bigl[
			\Delta^{+}_{\psi_{\pm}} \sgn(x^{-}-y^{-})
			- \Delta^{-}_{\psi_{\pm}}\sgn(x^{+}-y^{+})
		\bigr]}
	}{
		2\pi
		\left(
			\frac{v\beta_0}{\pi}
			\sinh \left( \frac{\pi}{v\beta_0} |x^{-}-y^{-}| \right)
		\right)^{2\Delta^{+}_{\psi_{\pm}}}
		\left(
			\frac{v\beta_0}{\pi}
			\sinh \left( \frac{\pi}{v\beta_0} |x^{+}-y^{+}| \right)
		\right)^{2\Delta^{-}_{\psi_{\pm}}}
	}.
\end{multline}  
Thus, it follows from \eqref{primneq} that the corresponding nonequilibrium correlation functions are
\begin{equation}
\bigl\langle J_{\pm}(x^\mp)J_{\pm}(y^\mp) \bigr\rangle^\infty_{\text{neq}}
= -\frac{K}{4 v^2\beta(x^\mp)\beta(y^\mp)
  	\sinh^2 \left( \int_{y^\mp}^{x^\mp} \frac{\pi}{v\beta(x')} dx' \right)}
\end{equation}
and
\begin{multline}
\hspace*{-0.2cm}\bigl\langle \psi_{\pm}(x^-,x^+)\psi^\dagger_{\pm}(y^-,y^+) 
\big\rangle^\infty_{\text{neq}} \\
{}\hspace*{-0.2cm}=	\frac{
		\ee^{\pi \ii \bigl[
			\Delta^{+}_{\psi_{\pm}}
				\sgn \bigl( \beta_0 \int_{y^{-}}^{x^{-}} \beta(x')^{-1} dx' \bigr)
			- \Delta^{-}_{\psi_{\pm}}
				\sgn \bigl( \beta_0 \int_{y^{+}}^{x^{+}} \beta(x')^{-1} dx' \bigr)
		\bigr]}
	}{
		2\pi\hspace{-0.05cm}
		\left(
			\frac{v\sqrt{\beta(x^-)\beta(y^-)}}{\pi}
			\sinh \bigl| \int_{y^{-}}^{x^{-}} \frac{\pi}{v\beta(x')}dx' \bigr|
		\right)^{2\Delta^{+}_{\psi_{\pm}}}\hspace{-0.15cm}
		\left(
			\frac{v\sqrt{\beta(x^+)\beta(y^+)}}{\pi}
			\sinh \bigl| \int_{y^{+}}^{x^{+}} \frac{\pi}{v\beta(x')}dx' \bigr|
		\right)^{2\Delta^{-}_{\psi_{\pm}}}
          }.\hspace*{-0.2cm}	
\end{multline}
We note that the latter agrees with Eq.\ (19) in \cite{LLMM2} to first order in the expansion parameter $\epsilon$ in \eqref{betaeps} and exactly reproduces Eq.\ (10) in \cite{LLMM2} in the long-time limit.


\subsection{Temperature and velocity profiles}
\label{subsec:4.2}

It is straightforward to generalize the argument of
Sect.~\ref{subsec:3.2} to nonequilibrium states as in \eqref{cO1} with
$\cG$ given by \eqref{cG2}, which is more conveniently written as
\begin{equation} 
\label{cGpm}
\cG = v\int_{-L/2}^{L/2} \left[ \beta_{+}(x)T_{+}(x)+\beta_{-}(x)T_{-}(x) \right] dx,
\quad  
\beta_{\pm}(x) = \beta(x) \left[ 1 \pm \nu(x)/v \right]. 
\end{equation}
Our argument above goes through as it stands but with $U(f)$ replaced by $U(f_+,f_-)= U_{+}(f_+)U_{-}(f_-)$ with two different diffeomorphisms $f_{\pm}$.
Choosing them as 
\begin{equation} 
\label{fpm}
f_{\pm}(x) = \int_{0}^x \frac{\beta_{0,\pm}}{\beta_{\pm}(x')} \, dx',
\quad
\frac{1}{\beta_{0,\pm}} = \frac{1}{L}\int_{-L/2}^{L/2}\frac1{\beta_{\pm}(x)} \, dx, 
\end{equation} 
one straightens out both profiles $1/\beta_{\pm}(x)$, replacing \eqref{transl} by the identity
\begin{equation}
\label{translnu}
\big\langle\cO\big\rangle_{\text{neq}}
= \big\langle U(f_+,f_-)\cO U(f_+,f_-)^{-1} \big\rangle_{\beta_{0,+},\beta_{0,-}},
\end{equation}
where
\begin{equation}
\label{GGE}  
\bigl\langle \cO\bigr\rangle_{\beta_{0,+},\beta_{0,-}}
= \frac{ \Tr\big(\ee^{-\beta_{0,+}H_{+} - \beta_{0,-}H_{-}}\cO\big)}
		{ \Tr\big(\ee^{-\beta_{0,+}H_{+} - \beta_{0,-}H_{-}}\big)},
\quad
H_{\pm}
= v \int_{-L/2}^{L/2} T_{\pm}(x) \, dx,
\end{equation}
define the expectations in a simple example of a generalized Gibbs state
with different temperatures for the right and the left movers.
In particular, \eqref{neqexp} becomes
\begin{equation}
\label{neqexpd} 
\Bigl\langle \prod_{j}T_{r_j}(x_j^{-r_j}) \Bigr\rangle_{\text{neq}}
= \Bigl\langle \prod_{j} \left( f'_{r_j}(x_j^{-r_j})^2\,T_{r_j}(f_{r_j}(x_j^{-r_j}))
- \frac{c}{24\pi}(Sf_{r_j})(x_j^{-r_j}) \right)
\Bigr\rangle_{\beta_{0,+},\beta_{0,-}}.  
\end{equation}
The thermodynamic limit of the expectations in \eqref{GGE} of the observables in \eqref{Tpms} may still be conveniently studied by going to the dual picture which, upon setting $\beta_{0,\pm} = \beta_0 (1 \pm {\nu_0}/{v})$, takes the form\footnote{This is proven using modular invariance for imaginary $\nu_0$ and continuing analytically to real $\nu_0$.} 
\begin{equation}
\label{dualtau}
\hspace*{-0.3cm}
\Bigl\langle \prod_{j} T_{r_j}(x_j) \Bigr\rangle_{\beta_{0,+},\beta_{0,-}} \! 
= \biggl\langle\hspace{-0.05cm}\cT \prod_{j}\hspace{-0.05cm}
        \left[
		- \frac{1}{{(1+r_j{\nu_0}/{v})^{2}}}
		T_{r_j}\hspace{-0.1cm}
                \left( \ii r_j \frac{{x_j}}{{(1+r_j{\nu_0}/{v})}} \right)
	\right]
	\biggr\rangle_{\hspace{-0.15cm}\frac{L}{v+\nu_0}, \frac{L}{v-\nu_0}}\hspace{-0.1cm},\hspace*{-0.3cm}
\end{equation}
where on the right-hand side the expectation is in the theory on the circle with circumference $v\beta_0$.
We infer that in the thermodynamic limit
\begin{align}
\Big\langle \prod_{j} T_{r_j}(x_j) \Big\rangle^\infty_{\beta_{0,+},\beta_{0,-}} \! 
& = \bigl\langle0\bigl|
			\cT \prod_{j} \left[
				- \frac{_1}{^{(1+r_j{\nu_0}/{v})^{2}}}
					T_{r_j} \left( \ii r_j\frac{_{x_j}}{^{(1+r_j{\nu_0}/{v})}} \right)
			\right]
		\bigr|0\bigr\rangle \nonumber \\
& = \bigl\langle0\bigl|
			\cT \prod_{j \, : \, r_j=+} \big( - T_{+}(\ii x_j) \big)
		\bigr|0\bigr\rangle
		\bigl\langle0\bigl|
			\cT \prod_{j \, : \, r_j=-} \big( - T_{-}(-\ii x_j) \big)
		\bigr|0\bigr\rangle,
\end{align}
where on the right-hand side the first (second) vacuum expectation is in the theory on the circle with circumference $v\beta_{0,+}$ ($v\beta_{0,-}$), and the last equality follows by the factorization in \eqref{factor} of the vacuum expectations and their rescaling.
In a similar way as in Sect.~\ref{subsec:3.3}, this shows that the identity in \eqref{neqexpd} holds in the thermodynamic limit with the infinite-volume profiles $\beta_{\pm}(x)>0$ with arbitrary positive asymptotic values $\beta_{\pm,\cL}$ and $\beta_{\pm,\cR}$ and $f_{\pm}(x)$ given by the first of the equations in \eqref{fpm} with arbitrary $\beta_{0,\pm}>0$.
In particular,
\begin{equation}
\big\langle T_{\pm}(x^{\mp})\big\rangle^\infty_{\text{neq}}
= \frac{\pi c}{12(v\beta_{\pm}(x^\mp))^2}-\frac{c}{24\pi}(Sf_{\pm})(x^\mp),
\end{equation}
implying that
\begin{equation} 
\label{EP2}
\bigl\langle\cE(x,t) \bigr\rangle^\infty_{\text{neq}}
= \frac{1}{2} \left[ F_{+}(x^-) + F_{-}(x^+) \right],
\quad 
\bigl\langle \cJ(x,t \bigr\rangle^\infty_{\text{neq}}
= \frac{v}{2} \left[ F_{+}(x^-) - F_{-}(x^+) \right] 
\end{equation}
with the functions
\begin{equation} 
\label{G2}
F_{\pm}(x)
= \frac{\pi c}{6v}\frac{1}{\beta_{\pm}(x)^2} + \frac{vc}{12\pi}
	\left(
		\frac{\beta_{\pm}''(x)}{\beta_{\pm}(x)} - \frac{1}{2}
		\left( \frac{\beta_{\pm}'(x)}{\beta_{\pm}(x)} \right)^2
	\right), 
\end{equation}
as described in Sect.~\ref{sec:Intro}.

There is a conceptual gain from the consideration of the nonequilibrium states with different temperature profiles for the right and left movers: unlike the states with equal profiles, such states are preserved by the Schr\"odinger-picture dynamics.
Indeed, for $\cG$ given by \eqref{cGpm},
\begin{equation}
\ee^{-\ii tH}\cG\ee^{\ii tH}
= v \int_{-L/2}^{L/2}
	\left[ \beta_{+}(x^-)T_{+}(x) + \beta_{-}(x^+) T_{-}(x) \right] dx
\end{equation}
so that under the Schr{\"o}dinger-picture evolution the profiles $\beta_{\pm}(x)$ move ballistically to the right and to the left, respectively.
This still holds in the thermodynamic limit and makes it clear why for long times such states converge to the generalized Gibbs state in \eqref{GGE} with inverse temperatures $\beta_{0,+} = \beta_{+,\cL}$ and $\beta_{0,-} = \beta_{-,\cR}$.
Conversely, we may view the nonequilibrium states with $\cG$ given by \eqref{cGpm} as the generalized Gibbs state with local profiles $\beta_{\pm}(x)$ whose
time evolution under the Schr\"odinger-picture dynamics reduces to the
time evolution of the profiles governed by the equations
\begin{equation}
\label{hydro_pic}
\partial_{t} \beta_{\pm} \pm v \partial_{x} \beta_{\pm} = 0.
\end{equation}
This is reminiscent of the time evolution of the generalized Gibbs states
with local profiles in the generalized hydrodynamic picture 
of integrable models out of equilibrium
\cite{BCNF,CBT,BVKM2,DSpY,DSp,DoYo,IlDeN,CDDKY,Do}. 
In CFT, however, no hydrodynamic-scale closure is needed to obtain 
the hydrodynamic evolution equation \eqref{hydro_pic} \cite{BeDo2}.


\subsection{Temperature and chemical-potential profiles}
\label{subsec:4.3}

Suppose that our CFT possesses a $\text{U(1)}$ symmetry generated by a conserved current with components $\rho$ and $j$ satisfying 
\begin{equation}
\partial_{t} \rho + \partial_{x} j = 0.
\end{equation}
For instance, $\rho = \rho_{+}+\rho_{-}$ and $j = Kv (\rho_{+}-\rho_{-})$ for
the local Luttinger model, see, e.g., Appendix~B in \cite{LLMM1} (we recall that $K$ is the 
Luttinger parameter).
One may then consider states with both temperature and chemical-potential\footnote{The name is somewhat conventional.
If $\rho$ is charge density, rather than the particle density, then $-\mu(x)$ would be the electric potential.} profiles by taking 
\begin{equation}
\label{cGmu}  
\cG = \int_{-L/2}^{L/2}\beta(x) \left[ \cE(x) - \mu(x)\rho(x) \right] dx,
\end{equation}
where $\rho(x)$ is the zero time density and $\mu(x)$ is a periodic chemical-potential profile.
Suppose that the light-cone components $J_{\pm} = (1/2)(\rho \pm v^{-1} j)$ of the current depend only on $x^\mp$, respectively, and satisfy the $\text{U(1)}$ current algebra:
\begin{align}
& [J_{\pm}(x),J_{\pm}(y)]
= \pm \frac{\kappa}{2\pi\ii}\delta'(x-y),
&& [J_{\pm}(x),J_{\mp}(y)]
= 0, \\ 
& [T_{\pm}(x),J_{\pm}(y)]
= \mp \ii\delta'(x-y)J_{\pm}(y) \pm \ii\delta(x-y)J'_{\pm}(y),
&& [T_{\pm}(x),J_{\mp}(y)]
= 0.
\end{align}
In terms of the Fourier modes,
\begin{equation}
J_{\pm}(x)
= \frac{1}{L} \sum_{n=-\infty}^\infty
	\ee^{\pm\frac{2\pi\ii n x}{L}}J^\pm_n
\end{equation}
and the commutation relations take the form
\begin{equation}
[J^\pm_n,J^\pm_m] = \kappa n\delta_{n+m,0},
\quad
[J^\pm_n,J^\mp_m] = 0,
\quad
[L_n^\pm,J^\pm_m] = -mJ^\pm_{n+m},
\quad
[L_n^\pm,J^\mp_m] = 0.
\end{equation}
For concreteness, we shall restrict our discussion to the case of the local Luttinger model in which case $J_{\pm}=\sqrt{K}\,\tilde{\rho}_{\pm}$ and $\kappa = K$, see \eqref{LutTpm}, or to the case when $J_{\pm}$ is one of the Cartan subalgebra components of the two current algebras of the level $k$ WZW theory based on a compact Lie group \cite{FMS} (e.g., $\text{U(1)}$ or $\text{SU(}N\text{)}$)
in which case $\kappa={k}/{2}$.
It follows from \cite{GW1,GW2} that in these cases there exist, besides the two projective representations $U_{\pm}$ of $\wDiff_+(\SL)$ considered above for which
\begin{equation}
U_{\pm}(f)J_{\pm}(x)U_{\pm}(f)^{-1} = f'(x)J_{\pm}(f(x)),
\quad
U_{\pm}(f)J_{\mp}(x)U_{\pm}(f)^{-1} = J_{\mp}(x),
\end{equation}
also two commuting projective representations $V_{\pm}$ of the additive gauge group of periodic smooth maps $h(x)=h(x+L)$ on $\mathbb R$ generated infinitesimally by $J_{\pm}$,
\begin{equation}
V_{\pm}(h)
= I \mp \ii\varepsilon\int_{-L/2}^{L/2}\xi(x)J_{\pm}(x) \, dx + o(\varepsilon)
\end{equation}
for $h(x) = \varepsilon\xi(x)$.
Under the adjoint action of $V_\pm(h)$,
\begin{align}
\hspace{-0.15cm}
V_{\pm}(h)J_{\pm}(x)V_{\pm}(h)^{-1}
& \! = J_{\pm}(x)\hspace{-0.04cm}+\hspace{-0.04cm}
  \frac{\kappa}{2\pi}h'(x),
& \hspace{-0.2cm} V_{\pm}(h)J_{\mp}(x)V_{\pm}(h)^{-1}
& \! = J_{\mp}(x), \\
\hspace{-0.15cm}
V_{\pm}(h)T_{\pm}(x)V_{\pm}(h)^{-1}
& \! = T_{\pm}(x)\hspace{-0.04cm} +\hspace{-0.04cm} h'(x)J_{\pm}(x)\hspace{-0.08cm} +\hspace{-0.08cm} \frac{\kappa}{4\pi}h'(x)^2,
& \hspace{-0.2cm} V_{\pm}(h)T_{\mp}(x)V_{\pm}(h)^{-1}
& \! = T_{\mp}(x).
\end{align}

Recall from Sect.~\ref{subsec:3.2} that the conjugation with operators $U(f)= U_{+}(f)U_{-}(f)$ for $f\in\wDiff(\SL)$ satisfying \eqref{f} with $\beta_0$ given by \eqref{beta0} was previously used to straighten out the periodic inverse-temperature profile $\beta(x)$ to a constant one given by $\beta_0$.
We shall keep the function $f$ as before and choose
\begin{equation}
h(x) = \frac{1}{v} \int_0^x \left( \mu(x') - \frac{\beta_0}{\beta(x')}\mu_0 \right) dx',
\quad
\mu_0 = \frac{1}{L} \int_{-L/2}^{L/2} \mu(x) \, dx.
\end{equation}
Then conjugating $\cG$ in \eqref{cGmu} first with $V(h)= V_+(h)V_-(h)$ and then with $U(f)$ straightens out the chemical-potential and inverse-temperature profiles to $\mu_0$ and $\beta_0$, respectively,
\begin{equation}
U(f)V(h)\cG V(h)^{-1}U(f)^{-1}
= \int_{-L/2}^{L/2}\beta_0 \left[ \cE(y) - \mu_0\rho(y) \right] dy + const,
\end{equation}
leading to the identity
\begin{equation}
\big\langle\cO\big\rangle_{\text{neq}}
= \big\langle U(f)V(h)\cO V(h)^{-1} U(f)^{-1}\big\rangle_{\beta_0,\mu_0}.
\end{equation}
The thermodynamic limit can be controlled as before using the dual representation of the equilibrium expectations (that involves now the theory on a circle of circumference $v\beta_0$ with twisted boundary conditions).
In the infinite volume, one can again treat profiles with arbitrary asymptotic values.
Using the fact that in the theories under consideration,
\begin{equation}
\big\langle T_{\pm}(x)\big\rangle^\infty_{\beta_0,\mu_0}
= \frac{\pi c}{12(v\beta_0)^2}+\frac{\kappa\mu_0^2}{4\pi v^2},
\quad
\big\langle J_{\pm}(x)\big\rangle^\infty_{\beta_0,\mu_0}
= \frac{\kappa\mu_0}{2\pi v}
\end{equation}
in the thermodynamic limit, one obtains the identities in \eqref{EP} with
\begin{equation}
\label{Fmu}  
F(x)
= \frac{\pi c}{6v} \frac{1}{\beta(x)^2}
	+ \frac{cv}{12\pi} \left(
			\frac{\beta''(x)}{\beta(x)}
			- \frac{1}{2} \left( \frac{\beta'(x)}{\beta(x)} \right)^2
		\right)
	+ \frac{\kappa}{2\pi v}\mu(x)^2
\end{equation}
as well as the formulas
\begin{equation}
\bigl\langle \rho(x,t) \bigr\rangle^\infty_{\text{neq}}
= \frac{1}{2} \left[ G(x^-) + G(x^+) \right],
\quad
\bigl\langle j(x,t) \bigr\rangle^\infty_{\text{neq}}
= \frac{v}{2} \left[ G(x^-) - G(x^+) \right]
\end{equation}
with
\begin{equation}
\label{Kmu}  
G(x) = \frac{\kappa}{\pi v} \mu(x).  
\end{equation}

For the matrix $\cD$ of the Drude weights obtained from the nonequilibrium expectations with respect to states with profiles with small kinks of heights $\Delta\beta=\beta_\cL-\beta_\cR$ and $\Delta\mu=\mu_\cL-\mu_\cR$ around the constant values $\beta_0$ and $\mu_0$ of the inverse temperature and chemical potential, the formula in \eqref{Drude} generalizes to
\begin{align}
\cD
& = \left( \begin{matrix}
		D_{11} & D_{12} \\
		D_{21} & D_{22}
	\end{matrix} \right) \nonumber \\
& = \lim_{\substack{\beta_{\cL,\cR}\to\beta_0 \\ \mu_{\cL,\cR}\to\mu_0}} \,
	\lim\limits_{t\to\infty}
        \frac{1}{t} \int
		\left( \begin{matrix}
		-\frac{\beta_0^2}{\Delta\beta} \big\langle \cJ(x,t) 
                \big\rangle^\infty_{\text{neq}}\Big|_{\Delta\mu=0}
		&  \frac{1}{\Delta\mu} \big\langle \cJ(x,t) 
                \big\rangle^\infty_{\text{neq}}\Big|_{\Delta\beta=0} \\
		-\frac{\beta_0^2}{\Delta\beta} 
                \big\langle j(x,t)\big\rangle^\infty_{\text{neq}}\Big|_{\Delta\mu=0}
		&  \frac{1}{\Delta\mu}\big\langle j(x,t) 
                \big\rangle^\infty_{\text{neq}}\Big|_{\Delta\beta=0}
	\end{matrix} \right) dx.
\label{Drudmat}
\end{align}
This yields
\begin{equation}
\cD
= \left( \begin{matrix}
	\frac{\pi vc}{3\beta_0}	&	\frac{\kappa v\mu_0}{\pi}\cr
	0												&	\frac{\kappa v}{\pi}
\end{matrix} \right) .
\end{equation}
Note that the thermal Drude weight $D_{11}$ is independent of $\mu_0$ and the density Drude weight $D_{22}$ is independent of the temperature.
In particular, for the Luttinger model, $D_{22} = {Kv}/{\pi}$.
The lack of symmetry of $\cD$ is due to the asymmetric way in which
the temperature and the chemical potential enter into the Gibbs state.\footnote{The coefficients 
of linear response of currents $\cJ$ and $j$ to small $-\Delta\beta$ and $\Delta(\beta\mu)$ 
would form a symmetric matrix that unveils a ballistic version of the Onsager reciprocal relations.}

As before, one may also consider nonequilibrium states with different local profiles $\beta_{\pm}(x)$ and $\mu_{\pm}(x)$ for right and left movers, which are defined by \eqref{cO1} with
\begin{equation}
\label{cG_most_general}
\cG
= \int_{-L/2}^{L/2} \sum_{r = \pm} \beta_{r}(x)
	\left[ v T_{r}(x) - \mu_{r}(x) J_{r}(x) \right] dx.
\end{equation}
This leads to the replacement of the functions $F$ and $G$ in \eqref{Fmu} and \eqref{Kmu} with
\begin{equation}
F_{\pm}(x)
= \frac{\pi c}{6v} \frac{1}{\beta_{\pm}(x)^2}
	+ \frac{cv}{12\pi} \left(
			\frac{\beta''_{\pm}(x)}{\beta_{\pm}(x)}
			- \frac{1}{2} \left( \frac{\beta'_{\pm}(x)}{\beta_{\pm}(x)} \right)^2
		\right)
	+ \frac{\kappa}{2\pi v}\mu_{\pm}(x)^2
\end{equation}
and
\begin{equation}
G_{\pm}(x) = \frac{\kappa}{\pi v} \mu_{\pm}(x),
\end{equation}
respectively.
In this case, the expectation values are given by \eqref{EP2} for $\bigl\langle \cE(x,t) \bigr\rangle^\infty_{\text{neq}}$ and 
$\bigl\langle \cJ(x,t) \bigr\rangle^\infty_{\text{neq}}$ as well as
\begin{equation}
\bigl\langle \rho(x,t) \bigr\rangle^\infty_{\text{neq}}
= \frac{1}{2} \left[ G_{+}(x^-) + G_{-}(x^+) \right],
\quad
\bigl\langle j(x,t) \bigr\rangle^\infty_{\text{neq}}
= \frac{v}{2} \left[ G_{+}(x^-) - G_{-}(x^+) \right].
\end{equation}
The nonequilibrium states in \eqref{cG_most_general} form the family of generalized Gibbs states with local profiles that correspond to the commuting conserved charges $H_{\pm}$ of \eqref{GGE} and $Q^{J}_{\pm}= \int_{-L/2}^{L/2}
J_{\pm}(x) \, dx$.
Such a family of states is again preserved by the
Schr\"odinger evolution
that displaces the local profiles $\beta_{\pm}(x)$ and $\mu_{\pm}(x)$ ballistically.
Their time evolutions are governed by \eqref{hydro_pic} together with
\begin{equation}
\partial_{t} \mu_{\pm} \pm v \partial_{x} \mu_{\pm} = 0.
\end{equation}

In the long-time limit, one obtains a genuine 
(i.e., with constant profiles) generalized Gibbs state 
which is the thermodynamic limit of the state with
\begin{equation}
\cG=\beta_{+,\cL}(H_+-\mu_{+,\cL}Q^J_+)+\beta_{-,\cR}(H_--\mu_{-,\cR}Q^J_-).
\end{equation}
In particular,
\begin{align}
\lim_{t\to\infty} \langle \cE(x,t) \rangle^\infty_{\text{neq}}
& = \frac{\pi c}{12v} \left( \beta_{+,\cL}^{-2}+\beta_{-,\cR}^{-2} \right)
		+ \frac{\kappa}{4\pi v}\left(\mu_{+,\cL}^2+\mu_{-,\cR}^2\right), \\
\lim_{t\to\infty} \langle \cJ(x,t) \rangle^\infty_{\text{neq}}
& = \frac{\pi c}{12} \left( \beta_{+,\cL}^{-2} - \beta_{-,\cR}^{-2} \right)
		+ \frac{\kappa}{4\pi}\left(\mu_{+,\cL}^2-\mu_{-,\cR}^2\right), \\
\lim_{t\to\infty} \langle \rho(x,t) \rangle^\infty_{\text{neq}}
& = \frac{\kappa}{2\pi v}\left(\mu_{+,\cL}+\mu_{-,\cR}\right),
		\label{permrho} \\
\lim_{t\to\infty} \langle j(x,t) \rangle^\infty_{\text{neq}}
& = \frac{\kappa}{2\pi}\left(\mu_{+,\cL}-\mu_{-,\cR}\right)
		\label{permj}
\end{align}
generalizing \eqref{15}. 

As a specific example, consider the Luttinger model with the initial state
given by  \eqref{cG_most_general} with $\beta_{\pm}(x) = \beta_{0}$ and with 
$\mu_{\pm}(x)$ interpolating when $L\to\infty$ between the asymptotic
values $\mu_{\pm,\cL}$ and $\mu_{\pm,\cR}$. The charges $Q^{J}_{\pm}=\sqrt{K}
\int\tilde\rho_\pm(x)dx$ are then
the number operators of right and 
left moving plasmons. One can also consider in that case the number operators of electrons (and holes) $Q^{e}_{\pm} =\int \rho_{\pm}(x) dx$, which are different from $Q^{J}_{\pm}$, although the total charges are equal, i.e., $Q^{e}_{+}
+ Q^{e}_{-} = Q^{J}_{+} + Q^{J}_{-}$. Indeed, we infer from \eqref{LutTpm} that
\begin{equation}
Q^J_\pm=\frac{K+1}{2}Q^{e}_\pm-\frac{K-1}{2}Q^{e}_\mp.
\end{equation}
Unlike for $Q^{J}_{\pm}$, the spectra of $Q^{e}_{\pm}$ are composed of integers.
In terms of $Q^{e}_{\pm}$, the generalized Gibbs state appearing in the long-time limit of the evolution will correspond to
\begin{equation}
\cG
= \beta_0(H -\mu^{e}_{+}Q^{e}_{+}- \mu^{e}_{-}Q^{e}_{-} ),
\quad
\mu^{e}_{\pm}
= \frac{1}{2} (\mu_{+,\cL} + \mu_{-,\cR})\pm \frac{K}{2} (\mu_{+,\cL} - \mu_{-,\cR}).
\end{equation}
From \eqref{permj} we infer that the value of the permanent 
current in the limiting nonequilibrium steady state is
\begin{equation}
\label{Imu}  
I
= \frac{K}{2\pi} \left( \mu_{+,\cL} - \mu_{-,\cR} \right)
= \frac{1}{2\pi} \left( \mu^{e}_{+} - \mu^{e}_{-} \right).
\end{equation}
It was argued in \cite{MS,ACF,Kaw} that $\mu^{e}_{\pm}$ that couple in the
steady state to the electron charges correspond
to the chemical potentials of free electrons of wide leads connected
to a Luttinger wire, at least if $\mu_+(x)=\mu_-(x)$. As was discussed in
\cite{LLMM1}, the second equality of \eqref{Imu} would then provide an
explanation for the experimental measurements \cite{THS} of conductance in
quantum wires that gave results close to the universal constant ${e^2}/{h}$,
equal to ${1}/{2\pi}$ in the units $\hbar = e = 1$ that we are using.
This universal value is different from the conductance ${K e^2}/{h}$ 
predicted in \cite{FK} which, instead, is consistent with
the first equality of \eqref{Imu} that uses the asymptotic values of the
imposed chemical-potential profiles that couple in the steady
state to the plasmon charges.  
The above extends the derivation of the universal result obtained in \cite{LLMM1} to states with constant temperature and with chemical-potential profiles 
possibly different for the right and left movers.\footnote{The result in \cite{LLMM1} was more general in that it was for the Luttinger model with nonlocal interactions, but it was only for zero temperature states and $\mu_+(x)=\mu_-(x)=\mu(x)$.}

The Luttinger model possesses also a conserved axial current 
with $\rho_A=\rho_+-\rho_-$ and 
$j_A=(v/K)(\rho_++\rho_-)$ satisfying $\partial_t\rho_A+\partial_xj_A=0$.
Note that \eqref{permrho}
implies that the permanent axial current in the limiting 
nonequilibrium steady state considered above takes the value
\begin{equation}
I_A
= \frac{1}{2\pi} \left( \mu_{+,\cL}+\mu_{-,\cR} \right)
= \frac{1}{2\pi} \left( \mu^{e}_++\mu^{e}_- \right)
\end{equation}
with the universal coefficient both when expressed in terms of the
asymptotic values of the profiles $\mu_\pm(x)$ and in terms of $\mu^{e}_\pm$.


\nsection{Equilibrium dynamics and relation to Euclidian CFT}
\label{sec:5}

In this section, we discuss the relation between
the articles \cite{DSVC,BD1,DSC,BD2} and the present paper.
In \cite{DSVC} it was argued that the kernel of the 1-particle density matrix
in the ground state of a nonrelativistic high density Fermi gas in a trap
may be described on mesoscopic scales by the 2-point function of the fermionic
massless free field whose Fermi velocity varies in space. These results
were generalized in \cite{BD1,DSC,BD2} to certain nonrelativistic systems
of interacting $1d$ bosons in traps. Despite similarities, there are several
differences with the approach of the present paper. First, the arguments
in \cite{DSVC} were based on the analysis of the ground-state Euclidian-time
correlators in the presence of a trap and these were shown to correspond
to Euclidian CFT correlators in an appropriate curved metric (in \cite{BD2}
also the coupling to a gauge field appeared implicitly). In this paper, we
consider positive temperatures, but the correspondence of \cite{DSVC}
generalizes to low-temperature states leading to the compactification of
the Euclidian time direction in CFT, just as for homogeneous equilibria.
Hence, for specific CFTs, the states in (\ref{cO1}) may, indeed, be viewed
as describing on mesoscopic scales the $1d$ nonrelativistic low-temperature
matter in traps, with $\beta(x)$ having the interpretation of the
position-dependent Fermi velocity in appropriate units. Second, the argument
of \cite{DSVC} was done for the equilibrium dynamics (although some
nonequilibrium situations were also considered), whereas in the bulk of
the present paper we study the dynamics generated by the homogeneous
Hamiltonian that does not preserve the states in (\ref{cO1}). Our considerations
may, however, be generalized to dynamics induced by inhomogeneous Hamiltonians
\cite{PM}, in particular to the ones that preserve the states in (\ref{cO1}).
Third, we use the Minkowski version of CFT, whereas the papers
\cite{DSVC,BD1,DSC,BD2} employed the Euclidian CFT. That is usually
considered as an innocent distinction handled by the Wick rotation.
Indeed, for the primary fields, the correlators in the Euclidian theory
in the metric considered in \cite{DSVC,BD1,DSC,BD2} and with compactified
time do agree, up to the Wick rotation, with the corresponding correlators
in the states of (\ref{cO1}) for which the time dependence is generated by
the equilibrium dynamics. As shown below, however, that does not hold directly
for the correlators of the energy-momentum tensor components which are of main
interest in this paper. This points to the need of caution when one applies
Euclidian techniques in the study of systems that are inhomogeneous
in space or/and time.
\vskip 0.1cm

To be more concrete, let us briefly discuss 
the equilibrium dynamics for the states in \eqref{cO1} that is generated by
the inhomogeneous Hamiltonian $\tilde H=\beta_0^{-1}\cG$. Defining
\begin{equation}
T_\pm(x;t)=\ee^{\ii t\tilde H}T_\pm(x)\,\ee^{-\ii t\tilde H}\,,\qquad
\Phi_j(x;t)=\ee^{\ii t\tilde H}\Phi_j(x,x)\,\ee^{-\ii t\tilde H}, 
\end{equation}
we immediately obtain from \eqref{transl} and \eqref{cond} the relations
\begin{align}
&\Big\langle \prod_{j} T_{r_j}(x_j;t_j) \Big\rangle_{\text{neq}}
=\Big\langle\prod_{j} \left( f'(x_j)^2T_{r_j}(f(x_j)^{-r_j}) -
\frac{c}{24\pi}(Sf)(x_j) \right)\Big\rangle_{\beta_0}\,,\label{Tcorreqdyn} \\
&\Big\langle \prod_{j}\Phi_j(x_j;t_j) \Big\rangle_{\text{neq}}\,=\Big\langle
\prod\limits_j\left(f'(x_j)^{\Delta^+_{\Phi_j}}f'(x_j)^{\Delta^-_{\Phi_j}}
\Phi_j(f(x_j)^-,f(x_j)^+)\right)\Big\rangle_{\beta_0},\label{Phicorreqdyn}
\end{align}
where $f(x_j)^{-r}=f(x_j)-rvt_j$. Note the difference of the right-hand sides
with those of \eqref{neqexp} and \eqref{primneq} corresponding to
the dynamics generated by the homogeneous Hamiltonian $H$ of \eqref{Hdef1}
which results in the time dependence in the arguments of function $f$. 
\vskip 0.1cm

After the Wick rotation $t_j=-i\tau_j$, the correlators in \eqref{Phicorreqdyn}
become the correlation functions of the same primary fields in the Euclidian
CFT on the torus $S^1\times S^1$ parameterized by $(x\ {\rm mod}\,L,\tau
\ {\rm mod}\,\beta_0)$ and equipped with the Riemannian metric
\begin{equation}
h=(dx)^2+(v\beta(x)/\beta_0)^2(d\tau)^2=\ee^{\sigma(z,\bar z)}h_0,
\label{metric}
\end{equation}
where $h_0=dzd\bar z$ for the complex coordinate $z=f(x)+\ii v\tau$
on the torus with $f(x)$ given by \eqref{f} and $\beta_0$ by \eqref{beta0},
and where $\sigma(z,\bar z)=-2\ln{f'(x)}$. In other words,
\begin{equation}
\Big\langle \prod_{j}\Phi_j(x_j,-i\tau_j)\Big\rangle_{\text{neq}}\,=\Big\langle
\prod_j\Phi_j(z_j,\bar z_j)\Big\rangle_{S^1\times S^1,\,h},
\label{WickEuc}
\end{equation}
where on the left-hand side is the Wick rotated \eqref{Phicorreqdyn}
and on the right-hand side the Euclidian correlation functions on
the torus $S^1\times S^1$ with the Riemannian metric $h$ of
\eqref{metric}. Such a relation is well known for $f(x)=x$. 
Its generalization to general $f$ follows from the identity
\begin{equation}
\Big\langle\prod_j\Phi_j(z_j,\bar z_j)\Big\rangle_{S^1\times S^1,\,\ee^\sigma h_0}
=\,\Big\langle
\prod_j\Big(e^{\frac{1}{2}(\Delta^+_{\Phi_j}+\Delta^-_{\Phi_j})
\sigma(z_j,\bar z_j)}\Phi_j(z_j,\bar z_j)\Big)\Big\rangle_{S^1\times S^1,\,h_0}.
\end{equation}
The ground-state version of the relations in \eqref{WickEuc} 
provided the basis for the use of Euclidian CFT in the description
of the trapped $1d$ fermions or bosons on mesoscopic scales
in \cite{DSVC,BD1,DSC,BD2}, with the interpretation of
$v\beta(x)/\beta_0$ as the position-dependent Fermi velocity clearly
reflected in the form of the metric in \eqref{metric}.
\vskip 0.1cm

Let us pass to the discussion of the energy-momentum correlators.
For $f(x)=x$, the Wick-rotated correlators in \eqref{Tcorreqdyn} are
represented by the Euclidian correlation functions of the energy-momentum
components $T_+=T_{zz}\Vert(dz)^2\Vert$ and
$T_-=T_{\bar z\bar z}\Vert(d\bar z)^2\Vert$ on the torus $S^1\times S^1$
with metric $h_0$. However, for general
$f(x)$ \,one has \cite{KG}, in the notation $z^+=z,z^-=\bar z$,  
\begin{align}
&\Big\langle\prod_jT_{r_j}(z_j,\bar z_j)\Big\rangle_{S^1\times S^1,\,\ee^\sigma h_0}\cr
&=\,\Big\langle
\prod_j\Big(e^{-\sigma(z_j,\bar z_j)}
\Big(T_{r_j}-\frac{c}{24\pi}\Big(\partial_{z^{r_j}}^2\sigma
-\frac{1}{2}(\partial_{z^{r_j}}\sigma)^2\Big)(z_j,\bar z_j)\Big)
\Big\rangle_{S^1\times S^1,\,h_0}\cr
&=\,\Big\langle
\prod_j\Big(f'(x_j)^2T_{r_j}(z_j,\bar z_j)+\frac{c}{48\pi}\Big(\frac{f'''}
{f'}-\Big(\frac{f''}{f'}\Big)^2\Big)
(x_j)\Big)\Big\rangle_{S^1\times S^1,\,h_0}
\end{align}
and the right-hand side does not represent correctly the Schwarzian-derivative
terms of the Wick-rotated \eqref{Tcorreqdyn}. In the Euclidian domain,
the Schwarzian derivative appears in the transformation law of the
energy-momentum components when one deals with holomorphic transformations
$z\mapsto f(z)$ \cite{FMS,KG}, but this is not the case here.
A closer examination shows that
\begin{equation}
\Big\langle\prod\limits_jT_{r_j}(x_j;-i\tau_j)
\Big\rangle_{\rm neq}
=\Big\langle\prod\limits_j\Big(-T_{r_j}(z_j,\bar z_j) - \frac{c}{48\pi}{\mathcal R}(z_j,\bar z_j)\Big) \Big\rangle_{S^1\times S^1,e^\sigma h_0},
\end{equation}
where ${\mathcal R}(z,\bar z)=-\beta''(x)/\beta(x)$ is the scalar curvature of the metric $e^\sigma h_0$.
This substantiates the comment made above about the need of caution.


\nsection{Conclusions}
\label{sec:6}


We elaborated on the formula of \cite{LLMM2} giving the full time evolution of the energy density and heat current from a nonequilibrium state with a preimposed temperature profile in the Luttinger model with local interactions. 
The formula was obtained in \cite{LLMM2} by expanding the nonequilibrium state around the equilibrium to all orders.
More details on the perturbative computation involving the exact calculation of complicated integrals, that may be interesting in its own right, were given.
The main part of the paper was devoted to showing how the formula of \cite{LLMM2}, a result of the resummation of the perturbative series, may be obtained using Minkowskian conformal symmetries of the local Luttinger model.
The idea was to use conformal transformations to map spatially inhomogeneous situations to homogeneous ones, straightening out a nonuniform temperature profile to a constant one.
This led to a direct relation between nonequilibrium and equilibrium states, 
yielding the remarkable formula of \cite{LLMM2} as a corollary.
The CFT argument holds for a general class of unitary CFTs and could be applied to
a wider class of nonequilibrium states that are preserved by the Schr\"odinger-picture evolution. The states in this class may be viewed as particular examples of 
simple generalized Gibbs states with local profiles, and they tend to ordinary
generalized Gibbs states at long times, 
somewhat similarly as in the scenario recently advocated for integrable 
models where the evolution at certain length and time scales could be 
described by generalized hydrodynamics
\cite{BCNF,CBT,BVKM2,DSpY,DSp,DoYo,IlDeN,CDDKY,Do}.
We obtained similar results also for CFTs with a $\text{U(1)}$ current 
algebra (including the local Luttinger model itself) where we treated 
nonequilibrium states with temperature and chemical-potential profiles.
Moreover, our results permit a more detailed analysis within CFT, compared 
to using the partitioning protocol studied before \cite{BeDo3}, of how 
a system starting in a state that looks like two different equilibria 
joint together evolves in time towards a nonequilibrium steady state 
described by a generalized Gibbs state. 

As was discussed in Sect.~\ref{sec:5}, at least some families
of the CFT nonequilibrium states that we studied in the present paper could 
be interpreted as providing a mesoscopic-scale description of dense
nonrelativistic $1d$ matter in macroscopic traps \cite{DSVC,BD1,DSC,BD2}.
The dynamical correlators in such CFT states should similarly describe
the corresponding nonrelativistic correlators at mesoscopic time scales
both for after-quench and for equilibrium dynamics. The other way of arriving
at the family of nonequilibrium CFT states that we considered is by reversing
the logic of this paper. In the periodic-volume Minkowski CFT, the conformal
symmetries (together with the gauge symmetries if a $\text{U(1)}$ current
algebra is present) are broken in the usual equilibria that are not preserved
by the symmetries. Instead, the application of the symmetry
transformations to the equilibrium states generates the family of
nonequilibrium states that were studied here.

In the infinite volume, the states with kink-like profiles 
give access to the full counting statistics of the 
energy or charge transfers through the kinks, similarly to the states 
arising in the partitioning protocol \cite{BeDo3}.
Although such statistics in both approaches differ at finite times, they 
have the same long-time large deviations.
This will be discussed elsewhere as it requires using different boundary 
conditions for finite volumes that allow one to avoid the duplication 
of kinks in the profiles.
Finally, another interesting exercise, which was abundantly discussed 
in the similar context of quantum quenches \cite{CaCa2}, concerns the 
evolution of the entanglement entropy or negativity starting from states 
with profiles of the type consider here.
By the replica trick, the latter may be extracted from nonequilibrium 
correlators of the twist primary fields in the replicated theory, to 
which our approach gives direct access.
The analysis of the corresponding formulas is left for future research.

\paragraph{Acknowledgements:}
All three of us profited from valuable input and encouragement by Joel Lebowitz
and Vieri Mastropietro. We gratefully acknowledge that this paper would
not have seen the light of day without them. We would also like to thank Jouko
Mickelsson and Herbert Spohn for helpful discussions. E.L.\ acknowledges
support by VR Grant No.\ 2016-05167. P.M.\ is thankful for financial support
from P.\ F.\ Lindstr{\"o}m's foundation (KTH travel scholarship VT-2017-0011
no.\ 4). The final version of the article has profited from pertinent
comments of the anonymous referees who, in particular, drew our attention to
the articles \cite{DSVC,BD1,DSC,BD2}.


\begin{appendix}


\section*{Appendix}
\label{Sec:Appendix}
\renewcommand{\theequation}{A.\arabic{equation}}

In this appendix we give a proof of Lemma~\ref{lemma:In}.

We find it convenient to write the integrals in \eqref{Jn'} as
\begin{equation}
\label{Jn1}
I_n(q_1,\ldots,q_n) = \frac{2}{v\bar\beta} \int_{\R}
\sum_{\nu \in (2\pi/\bar\beta)\Z} \prod_{j=0}^n f_\nu(p+Q_j) \, dp,
\quad
f_\nu(p) = \frac{vp}{\ii\nu-vp}.
\end{equation} 
To compute these integrals we insert the Taylor series 
\begin{equation} 
\label{Taylor}
f_\nu(p+Q_j)
= f_\nu(p) + f^{(1)}(p)Q_j + \frac{1}{2} f^{(2)}(p)Q_j^2 + \ldots,
\quad
f_\nu^{(j\geq 1)}(p)
= \frac{j! v^{j} \ii\nu}{(\ii\nu-v p)^{j+1}}
\end{equation} 
into the integrand and obtain 
\begin{equation}
\label{Jn2} 
I_n(q_1,\ldots,q_n)
= I_n^{(n+1)}
	+ I_n^{(n,1)} \sum_{j=0}^nQ_j
	+ \frac{1}{2} I_n^{(n,0,1)} \sum_{j=0}^nQ_j^2
	+ I_n^{(n,2)} \sum_{0\leq j<k\leq n}Q_jQ_k
	+ R_n
\end{equation} 
with the integrals 
\begin{equation} 
\label{JnB} 
I_n^{(m_0,m_1,\ldots,m_k)}
= \frac{2}{v\bar\beta} \int_{\mathbb{R}}
	\sum_{\nu \in (2\pi/\bar\beta)\Z}
	f_\nu(p)^{m_0}f^{(1)}_\nu(p)^{m_1}\ldots f^{(k)}_\nu(p)^{m_k} \, dp,
\end{equation} 
where $m_j = 1,2,\ldots$ for $j=0,1,\ldots,k$ such that $\sum_{j} m_j = n+1$,
and $R_n$ a linear combination of terms
\begin{equation} 
\label{Rest} 
I_n^{(m_0,m_1,\ldots,m_k)}
\prod_{j=1}^{k} Q_{\ell_{j,1}}^{j} \ldots Q_{\ell_{j,m_j}}^{j} 
\end{equation} 
with indices $1\leq \ell_{j,1} < \ldots < \ell_{j,m_j}\leq n$ for $j = 1,2,\ldots,k$ and $\{ m_j \}$ such that $\sum_j jm_j\geq 3$.

To compute the integrals in \eqref{JnB} we define $M=\sum_j jm_j$ and insert the derivatives of $f_\nu(p)$ from \eqref{Taylor}. This gives
\begin{multline} 
\label{Immm}
I_n^{(m_0,m_1,\ldots,m_k)} \\
\begin{aligned}
& = 2!\cdots k! \int_{\mathbb{R}} 2 v^{M+m_0-1} \frac{1}{\bar\beta}
		\sum_{\nu \in (2\pi/\bar\beta)\Z} \frac{p^{m_0}
		(\ii\nu)^{n-m_0+1}}{(\ii\nu-vp)^{n+M+1} } \, dp \\
& = 2! \ldots k! \int_{\mathbb{R}} \sum_{\ell}^{n-m_0+1}
		\binom{n-m_0+1}{\ell} 2 v^{M+m_0+\ell-1}
		\frac{1}{\bar\beta} \sum_{\nu \in (2\pi/\bar\beta)\Z}
		\frac{p^{m_0+\ell}}{(\ii\nu-vp)^{M+m_0+\ell}} \, dp \\
& = 2! \ldots k! \int_{\mathbb{R}} \sum_{\ell=0}^{n-m_0+1} \binom{n-m_0+1}{\ell}
		\frac{(v\bar\beta)^{M-2}s^{m_0+\ell} }{(M+m_0+\ell-1)!}
		\frac{d^{M+m_0+\ell-1}}{ds^{M+m_0+\ell-1}} \left( \coth(\half s) \right) ds.
\end{aligned}
\end{multline} 
In the second equality we wrote $(\ii\nu)^{n-m_0+1} = (\ii\nu-vp+vp)^{n-m_0+1}$ to expand into a binomial series, and in the third we summed the bosonic Matsubara frequencies $\nu$ using the Mittag-Leffler series of $\coth(v\bar\beta p/2)$ and changed variables to $s = v\bar\beta p$.
We note that, for $M\geq 1$, the integrand of the last integral is singular, but the singularity is removable, i.e., one can replace $\coth(s/2)$ by $ \coth(s/2)-2/s$ without changing the result.
To further compute these integrals we use partial integrations.
We find that the integrals in \eqref{Immm} are zero for $M \geq 3$, which implies $R_n = 0$.
The remaining integrals $I_n^{(n+1)}$, $I^{(n,1)}_n$, $I_n^{(n,0,1)}$, and $I_n^{(n-1,2)}$ are found by straightforward computations.  
Inserting them into \eqref{Jn2} we obtain the result in Lemma~\ref{lemma:In}. 


\end{appendix}




\begin{thebibliography}{99}
 
\bibitem{RLL} Z.\ Rieder, J.\ L.\ Lebowitz, and E.\ Lieb:
Properties of a harmonic crystal in a stationary nonequilibrium state.
J.\ Math.\ Phys.\ {\bf 8}, 1073 (1967)

\bibitem{SpLe} H.\ Spohn and J.\ L.\ Lebowitz:
Stationary non-equilibrium states of infinite harmonic systems.
Commun.\ Math.\ Phys.\ {\bf 54}, 97 (1977)

\bibitem{ZNP} X.\ Zotos, F.\ Naef, and P.\ Prelovsek:
Transport and conservation laws.
Phys.\ Rev.\ B {\bf55}, 11029 (1997)

\bibitem{HoAr} T.\ G.\ Ho and H.\ Araki:
Asymptotic time evolution of a partitioned infinite two-sided isotropic $XY$-chain.
Tr.\ Mat.\ Inst.\ Steklova {\bf 228}, 203 (2000)

\bibitem{Og1} Y.\ Ogata:
Nonequilibrium properties in the transverse $XX$ chain.
Phys.\ Rev.\ E {\bf 66}, 016135 (2002) 

\bibitem{AsPi} W.\ H.\ Aschbacher and C.-A.\ Pillet:
Non-equilibrium steady states of the $XY$ chain.
J.\ Stat.\ Phys.\ {\bf 112}, 1153 (2003)

\bibitem{Gia} T.\ Giamarchi:
Quantum Physics in One Dimension. 
Oxford University Press (2004)

\bibitem{Zo} X.\ Zotos:
Issues on the transport of one dimensional systems. 
J.\ Phys.\ Soc.\ Jpn,\ Suppl. {\bf74}, 173 (2005) 

\bibitem{SPA1}
J.\ Sirker, R.\ G.\ Pereira, and I.\ Affleck:   
Diffusion and ballistic transport in one-dimensional
quantum systems. Phys.\ Rev.\ Lett.\ {\bf 103}, 216602 (2009)

\bibitem{SPA2} J.\ Sirker, R.\ G.\ Pereira, and I.\ Affleck:
Conservation laws, integrability, and transport in one-dimensional
quantum systems. Phys. Rev.\ B\ {\bf 83},\ 035115 (2011)
  
\bibitem{BDZ} I.\ Bloch, J.\ Dalibard, and W.\ Zwerger:
Many-body physics with ultracold gases.
Rev.\ Mod.\ Phys.\ {\bf 80}, 885 (2008)

\bibitem{PSSV} A.\ Polkovnikov, K.\ Sengupta, A.\ Silva, and M.\ Vengalattore:
Colloquium: Nonequilibrium dynamics of closed interacting quantum systems.
Rev.\ Mod.\ Phys.\ {\bf 83}, 863 (2011)

\bibitem{CaCh} M.\ A.\ Cazalilla and M.-C.\ Chung:
Quantum quenches in the Luttinger model and its close relatives.
J.\ Stat.\ Mech.\ 064004 (2016)

\bibitem{BCNF} B.\ Bertini, M.\ Collura, J.\ De Nardis, and M.\ Fagotti: 
Transport in out-of-equilibrium $XXZ$ chains: Exact profiles of charges
and currents. Phys.\ Rev.\ Lett.\ {\bf 117}, 207201 (2016)

\bibitem{CBT} O.\ A.\ Castro-Alvaredo, B.\ Doyon, and T.\ Yoshimura:
Emergent hydrodynamics in integrable quantum systems out of equilibrium. 
Phys.\ Rev.\ X {\bf 6}, 041065 (2016)

\bibitem{BVKM2} V.\ B.\ Bulchandani, R.\ Vasseur, C.\ Karrasch, and J.\ E.\ Moore:
Bethe-Boltzmann hydrodynamics and spin transport in the XXZ chain.
arXiv:1702.06146 [cond-mat.stat-mech] (2017)

\bibitem{DSpY} B.\ Doyon, H.\ Spohn, and T.\ Yoshimura:
A geometric viewpoint on generalized hydrodynamics.
arXiv:1704.04409 [cond-mat.stat-mech] (2017)

\bibitem{DSp} B.\ Doyon and H.\ Spohn:
Drude weight for the Lieb-Liniger Bose gas.
arXiv:1705.08141 [cond-mat.stat-mech] (2017)

\bibitem{Spo} H.\ Spohn:
Interacting and noninteracting integrable systems.
arXiv:1707.02159 [cond-mat.stat-mech] (2017)

\bibitem{DoYo} B. Doyon and T. Yoshimura:
A note on generalized hydrodynamics:
inhomogeneous fields and other concepts.
SciPost Phys. {\bf2}, 014 (2017)

\bibitem{IlDeN} E. Ilievski and J. De Nardis:
Ballistic transport in the one-dimensional Hubbard model: The hydrodynamic
approach. Phys. Rev. B {\bf96}, 081118(R) (2017) 

\bibitem{CDDKY} J.-S. Caux, B. Doyon, J. Dubail, R. Konik and T. Yoshimura:
Hydrodynamics of the interacting Bose gas in the Quantum Newton Cradle setup.
arXiv:1711.00873 [cond-mat.stat-mech] (2017)

\bibitem{Do} B. Doyon:
Exact large-scale correlations in integrable systems out of equilibrium.
arXiv:1711.04568 [math-ph] (2017)

\bibitem{CaCa1} P. Calabrese and J. Cardy:
Time dependence of correlation functions following a quantum quench.
Phys.\ Rev.\ Lett.\ {\bf 96}, 136801 (2006)

\bibitem{CaCa2} P.\ Calabrese and J.\ Cardy:
Quantum quenches in 1+1 dimensional conformal field theories.
J.\ Stat.\ Mech.\ 064003\ (2016)

\bibitem{BeDo1} D.\ Bernard and B.\ Doyon:
Energy flow in non-equilibrium conformal field theory.
J. Phys. A: Math. Theor. {\bf 45}, 362001 (2012) 

\bibitem{BeDo3} D.\ Bernard and B.\ Doyon:
Conformal field theory out of equilibrium: A review.
J.\ Stat.\ Mech.\ 064005\ (2016)

\bibitem{HL} S.\ Hollands and R. Longo: Non-equilibrium thermodynamics
and conformal field theory. Commun. Math. Phys. {\bf 357}, 43 (2018)
  
\bibitem{DSVC} J. Dubail, J.-M. St\'ephan, J. Viti and P. Calabrese:
Conformal field theory for inhomogeneous one-dimensional quantum systems:
the example of non-interacting Fermi gases. SciPost Phys. {\bf2}, 002 (2017)

\bibitem{BD1} Y. Brun and J. Dubail: One-particle density matrix of trapped
one-dimensional impenetrable bosons from conformal invariance.
SciPost Phys. {\bf2}, 012 (2017)

\bibitem{DSC} J. Dubail, J.-M. St\'ephan and P. Calabrese: Emergence
of curved light-cones in a class of inhomogeneous Luttinger liquids.
SciPost Phys. {\bf3}, 019 (2017)

\bibitem{BD2} Y. Brun and J. Dubail: The Inhomogeneous Gaussian Free Field,
with application to ground state correlations of trapped 1d Bose gases.
arXiv:1712.05262 [cond-mat.stat-mech] (2017)

\bibitem{LLMM2} E.\ Langmann, J.\ L.\ Lebowitz, V.\ Mastropietro, 
and P.\ Moosavi:
Time evolution of the Luttinger model with nonuniform temperature profile.
Phys.\ Rev.\ B {\bf 95},\ 235142\ (2017)

\bibitem{SoCa} S.\ Sotiriadis and J.\ Cardy:
Inhomogeneous quantum quenches. J.\ Stat.\ Mech.\ P11003\ (2008) 

\bibitem{ViRi} L.\ Vidmar and M.\ Rigol:
Generalized Gibbs ensemble in integrable lattice models.
J.\ Stat.\ Mech.\ 064007\ (2016) 

\bibitem{LLMM1} E.\ Langmann, J.\ L.\ Lebowitz, V.\ Mastropietro, 
and P.\ Moosavi:
Steady states and universal conductance in a quenched Luttinger model. 
Commun.\ Math.\ Phys.\ {\bf 349}, 551 (2017)

\bibitem{ML} D.\ C.\ Mattis and E.\ H.\ Lieb:
Exact solution of a many-fermion system and its associated boson field.
J.\ Math.\ Phys.\ {\bf 6}, 304 (1965)

\bibitem{LaMo} E.\ Langmann and P.\ Moosavi:
Construction by bosonization of a fermion-phonon model.
J.\ Math.\ Phys.\ {\bf 56}, 091902 (2015)

\bibitem{Voit} J.\ Voit:
One-dimensional Fermi liquids.
Rep. Prog. Phys. {\bf 58}, 977 (1995)

\bibitem{FMS} P.\ Di Francesco, P.\ Mathieu, and D.\ S{\'e}n{\'e}chal: 
Conformal Field Theory.
Springer, Berlin (1997)

\bibitem{Callen} H.\ N.\ Callen:
Thermodynamics and an Introduction to Thermostatics.
John Wiley \& Sons, New York (1985)

\bibitem{BeDo4} D.\ Bernard and B.\ Doyon:
Diffusion and signatures of localization in stochastic conformal field theory.
Phys.\ Rev.\ Lett.\ {\bf 119}, 110201 (2017)

\bibitem{MW} V.\ Mastropietro and Z.\ Wang:
Quantum quench for inhomogeneous states in the nonlocal Luttinger model. 
Phys.\ Rev.\ B {\bf 91}, 085123 (2015)

\bibitem{R} S.\ N.\ M.\ Ruijsenaars:
On Bogoliubov transformations for systems of relativistic charged particles.
J.\ Math.\ Phys.\ {\bf 18}, 517 (1977)

\bibitem{R1} S.\ N.\ M.\ Ruijsenaars:
On Bogoliubov transformations. II. The General Case.
Ann.\ Phys.\ {\bf 116}, 105 (1978)

\bibitem{GL} H.\ Grosse and E.\ Langmann:
A superversion of quasifree second quantization. I. Charged particles.
J.\ Math.\ Phys.\ {\bf 33}, 1032 (1992)

\bibitem{GW2} R.\ Goodman and N.\ R.\ Wallach:
Projective unitary positive-energy representations of $Diff(S^1)$.
J.\ Func.\ Anal.\ {\bf 63}, 299 (1985) 

\bibitem{VKM} R. Vasseur, C. Karrasch, and J. E. Moore: Expansion potentials
for exact far-from-equilibrium spreading of particles and energy.
Phys. Rev. Lett. {\bf115}, 267201 (2015)

\bibitem{PM} E. Langmann and P. Moosavi: in preparation

\bibitem{KG} K. Gaw\c{e}dzki: Lectures on Conformal Field Theory. In: Quantum
Fields and Strings: A Course for Mathematicians, Eds. P. Deligne {\it et al.}, 
AMS-IAS (1999), pp. 727-805

\bibitem{BeDo2} D.\ Bernard and B.\ Doyon: A hydrodynamic approach to 
non-equilibrium conformal field theories. J. Stat. Mech. 033104 (2016)

\bibitem{GW1} R.\ Goodman and N.\ R.\ Wallach:
Structure and unitary cocycle representations of loop groups and the group
of diffeomorphisms of the circle.
J.\ Reine Angew.\ Math.\ {\bf 347}, 69 (1984)

\bibitem{MS} D.\ L.\ Maslov and M.\ Stone:
Landauer conductance of Luttinger liquids with leads.
Phys. Rev. B {\bf 52}, R5539 (1995)

\bibitem{ACF} A.\ Y.\ Alekseev, V.\ V.\ Cheianov, and J.\ Fr{\"o}hlich:
Comparing conductance quantization in quantum wires and quantum Hall systems.
Phys.\ Rev.\ B {\bf 54}, R17320 (1996)

\bibitem{Kaw} A.\ Kawabata:
On the renormalization of conductance in Tomonaga-Luttinger liquid.
J.\ Phys.\ Soc.\ Jpn.\ {\bf 65}, 30 (1996)

\bibitem{THS} S.\ Tarucha, T.\ Honda, and T.\ Saku:
Reduction of quantized conductance at low temperatures observed in 2 to 10 $\mu$m-long quantum wires.
Sol.\ State Commun.\ {\bf 94}, 413 (1995) 

\bibitem{FK} C.\ L.\ Kane and M.\ P.\ A.\ Fisher:
Transport in a one-channel Luttinger liquid.
Phys.\ Rev.\ Lett.\ {\bf 68}, 1220 (1992)

\end{thebibliography}
\end{document}